\documentclass[11pt]{article}

\usepackage[normalem]{ulem}
\usepackage{longtable}
\usepackage{makeidx}
\usepackage{fullpage}
\usepackage{amsmath}
\usepackage{amssymb}
\usepackage{setspace}
\usepackage{bbm}
\usepackage{dsfont}
\usepackage{epsfig}
\usepackage[font=footnotesize,labelfont=bf,justification=centerlast,width=.94\textwidth]{caption}
\usepackage{cite}
 \usepackage{multirow}
\usepackage{array,booktabs}
\usepackage[cal=boondox]{mathalfa}
\usepackage{color}

\usepackage{hyperref}

\hypersetup{
    bookmarks=true,%
    colorlinks,%
    citecolor=blue,%
    filecolor=black,%
    linkcolor=black,%
    urlcolor=blue
}

\newcommand{\be}{\begin{equation}}
\newcommand{\ee}{\end{equation}}
\newcommand{\bes}{\begin{equation*}}
\newcommand{\ees}{\end{equation*}}
\newcommand{\bea}{\begin{eqnarray}}
\newcommand{\eea}{\end{eqnarray}}
\newcommand{\beas}{\begin{eqnarray*}}
\newcommand{\eeas}{\end{eqnarray*}}

\newcommand{\p}{\partial}

\newcommand\bell{\overline{\ell}}
\newcommand\bG{\overline{G}}
\newcommand{\bA}{\bar{A}}

\newcommand{\bmat}{\begin{bmatrix}}
\newcommand{\emat}{\end{bmatrix}}

\newcommand{\slt}{SL(2,\RR)}
\newcommand{\aslt}{sl(2,\RR)}

\newcommand{\RR}{\mathbb{R}}

\newcommand{\ZZ}{\mathbb{Z}}
\def\tr{{\rm tr}}
\def\Tr{{\rm Tr}}
\def\le{\left}
\def\ri{\right}
\def\ep{{\epsilon}}
\newcommand\sO{{\ensuremath{{\mathcal O}}}}

\newcommand{\al}{\alpha}

\def\Tr{{\rm Tr}}
\def\le{\left}
\def\ri{\right}

\def\CR{{\cal R}}

\newcommand{\R}{\mathbb{R}}

\newcommand{\sP}{\mathcal{P}}

\newcommand\sR{{\ensuremath{{\mathcal R}}}}
\newcommand{\Sig}{\Sigma}
\newcommand{\sig}{\sigma}
\newcommand{\sG}{\mathcal{G}}
\newcommand{\ha}{\frac{1}{2}}
\newcommand{\ben}{\begin{enumerate}}
\newcommand{\een}{\end{enumerate}}
\newcommand{\bz}{\bar{z}}
\newcommand{\bw}{\bar{w}}

\begin{document}
\numberwithin{equation}{section}
{
\begin{titlepage}
\begin{center}

\hfill \\
\hfill \\
\vskip 0.75in

{\Large \bf Wilson Lines and Ishibashi States in AdS$_3$/CFT$_2$}\\

\vskip 0.4in

{\large Alejandra Castro${}^{a}$, Nabil Iqbal${}^{b}$, and Eva Llabr\'es${}^{a}$}\\

\vskip 0.3in

${}^{a}${\it Institute for Theoretical Physics Amsterdam and Delta Institute for Theoretical Physics, University of Amsterdam, Science Park 904, 1098 XH Amsterdam, The Netherlands} \vskip .5mm
${}^{b}${\it Centre for Particle Theory, Department of Mathematical Sciences, Durham University,
South Road, Durham DH1 3LE, UK} \vskip .5mm

\texttt{a.castro@uva.nl, nabil.iqbal@durham.ac.uk, e.m.llabres@uva.nl}

\end{center}

\vskip 0.35in

\begin{center} {\bf ABSTRACT } \end{center}
We provide a refined interpretation of a gravitational Wilson line in AdS$_3$  in terms of Ishibashi states in the dual CFT$_2$.  Our strategy is to give a method to evaluate the Wilson line that accounts for all the information contained in the representation, and clarify the role of boundary conditions at the endpoints of the line operator. This gives a novel way to explore and reconstruct the local bulk dynamics which we discuss. We also compare our findings with other interpretations of Ishibashi states in AdS$_3$/CFT$_2$. 

\vfill

\noindent \today

\end{titlepage}
}

\newpage

\tableofcontents

\section{Introduction}
In this paper we discuss 3d gravity with negative cosmological constant. This is a topological theory with no local degrees of freedom. This fact can be made manifest by rewriting 3d gravity as a Chern-Simons theory with gauge group $SO(2,2) = \slt \times \slt$ \cite{Achucarro:1987vz,Witten:1988hc}. The Chern-Simons formulation has many advantages: BTZ black holes appear very naturally as topological defects around which the $\aslt$ gauge fields have non-trivial holonomies. Boundary gravitons can be understood as the usual edge excitations that appear when Chern-Simons theories are formulated on manifolds with boundary \cite{Henneaux:1999ib}. Diffeomorphisms may be easily understood on-shell as gauge transformations \cite{Witten:1988hc}. Finally, the Chern-Simons formulation is also very convenient for the extension to theories of higher spin gravity \cite{Aragone:1983sz,Blencowe:1988gj,Bergshoeff:1989ns,Henneaux:2010xg,Campoleoni:2010zq}. 

While the Chern-Simons formulation makes manifest the topological character of 3d gravity, it does so at a cost, by greatly obscuring geometric aspects. Simple geometric concepts such as proper distances or volumes are not at all transparent in the Chern-Simons formulation. The problem becomes even more acute if we consider coupling matter -- e.g. a simple scalar field -- to 3d gravity: this is very difficult to do in the Chern-Simons formulation, presumably because the theory is no longer purely topological (see however \cite{Grignani:1991nj} for some previous work in this direction). These facts make it very difficult to probe local bulk physics in the Chern-Simons formulation. This appears to be related to the fact that typical bulk observables such as (e.g.) the bulk-to-bulk propagator of a probe scalar field are not actually invariant under diffeomorphisms, and thus are difficult to formulate in a suitably gauge-invariant manner in Chern-Simons theory.   

Nevertheless, in AdS$_3$ the presence of a boundary allows the formulation of suitably diffeomorphism invariant observables -- the correlation functions of the dual CFT$_2$ -- and thus one would expect that it would be possible to compute such objects in the Chern-Simons formulation. Some progress in this direction was made in \cite{Ammon:2013hba,deBoer:2013vca,Castro:2014mza,deBoer:2014sna,Hegde2016}, motivated largely by the computation of entanglement entropy of field theories dual to 3d bulk higher spin gravity. In this work we will develop further the approach initiated by \cite{Ammon:2013hba}, where it was argued that a Wilson line in an infinite-dimensional highest-weight representation $\sR$ under the bulk $\slt \times \slt$ gauge group could be used to compute boundary theory correlators, i.e.:
\be\label{eq:wcr}
W_{\sR}(x_i,x_f) \underset{r\to \infty}{=} \langle \Psi| \sO (y_i) \sO(y_f)|\Psi \rangle~,
\ee
where we have picked coordinates $x^{\mu} = (r, y^i)$ with $r$ an AdS holographic coordinate and $y^i$ a CFT coordinate. Here the Wilson line $W_{\sR}$ ends on the boundary at $r \to \infty$, and $\Psi$ denotes the CFT$_2$ state dual to a particular configuration of Chern-Simons gauge fields that constitute the gravitational background in the interior. 

The representation space $\sR$ was generated from the Hilbert space of an auxiliary $\slt$-valued quantum mechanical degree of freedom $U(s)$ that lives on the Wilson line. The quadratic Casimirs of the representation $\sR$ mapped in the usual manner to the conformal dimensions $(h, \bar{h})$ of the dual CFT operator. While this represented progress towards extracting geometric observables from the Chern-Simons formulation of 3d gravity, several issues remained obscure: 
\ben
\item The relation \eqref{eq:wcr} was understood to hold only if a particular boundary condition was used for the auxiliary field $U$, demanding that it approached the identity element of $\slt$ at the two endpoints of the Wilson line. While this is perhaps a somewhat natural choice, its precise interpretation in the CFT was not made clear.  
\item All previous treatment of the $U(s)$ path integral was performed in a semi-classical limit, i.e. one in which $h \gg 1$. At a calculational level this allowed the path integral to be evaluated using its saddle-point; nevertheless this restriction seems somewhat artificial from the point of view of the dual CFT. Is it possible to go away from this limit? 
\item How can one obtain other bulk observables from the Chern-Simons formulation, e.g. bulk-to-bulk propagators or one-loop determinants for scalar fields on the gravitational background?
\een

In this work we answer these questions by providing a careful and fully quantum mechanical treatment of the Wilson line described above. In particular, we will show that the $U(s)$ worldline degree of freedom originally introduced in \cite{Ammon:2013hba} can be understood as a particular $\slt$ rotation of the global part of an Ishibashi state (familiar from boundary CFT). We use this technology to develop a purely algebraic method for computing open-ended Wilson lines, and demonstrate equivalence (in the semi-classical limit) with the path-integral techniques used in \cite{Ammon:2013hba}. 


%
The outline of this paper is as follows. In Sec. \ref{sec:path} we review the path integral  representation of  $W_{\sR}(x_i,x_f) $ proposed in \cite{Ammon:2013hba}, which will serve as a comparison to our quantum mechanical analysis. In Sec. \ref{sec:hilbert} we turn to a detailed analysis of the quantum mechanics responsible of the geometrical features in $W_{\sR}(x_i,x_f)$. This motivates the introduction of coherent states which we denote as \emph{rotated Ishibashi states}. Using these states, we relate  $W_{\sR}(x_i,x_f)$ to their inner product; we rederive the path integral formulation by discretizing this inner product; and we show that  $W_{\sR}(x_i,x_f)$  is a Green's function on the group manifold $SO(2,2)$. In Sec. \ref{sec:examples} we tie the quantum mechanical aspects of $W_{\sR}(x_i,x_f)$ to its geometrical features. We show that $W_{\sR}(x_i,x_f)$ is a Green's function on spacetime created by the Chern-Simons connections (which is a distinct statement from the properties on the group manifold). For global AdS$_3$ and the BTZ black hole, we show how to build local bulk fields by a suitable decomposition of  $W_{\sR}(x_i,x_f)$. This provides a new local probe of AdS$_3$ in the Chern-Simons formulation of 3d gravity. In Sec. \ref{sec:cft} we discuss the CFT interpretation of our results. And in Sec. \ref{sec:discussion} we discuss future directions and related results in AdS/CFT that make use of Ishibashi states.

\section{Path integral representation}\label{sec:path}

In this section we will consider the path integral representation of a Wilson line operator in the Chern-Simons theory. As we review below, this object should be thought of as the Chern-Simons description of the worldline of a massive particle moving in the bulk. This section is a brief summary of the results in \cite{Ammon:2013hba}. 

The gauge group of the Chern-Simons theory is $SO(2,2)\simeq SL(2,\mathbb{R})\times SL(2,\mathbb{R})$, and the bulk $sl(2,\mathbb{R})$ gauge connections are $A$, $\bar{A}$.\footnote{In appendix \ref{app:cs} we present our conventions on the Chern-Simons description of AdS$_3$ gravity.} The natural observables in Chern-Simons theory are Wilson loops in a certain representation $\sR$ of the bulk gauge group; in this work we will always take $\sR$ to be a product of two infinite-dimensional highest-weight representations in  $sl(2,\mathbb{R}) \oplus sl(2, \mathbb{R})$. 

We may now consider the following Wilson loop operator:
\begin{equation}\label{Wilsonfunc}
 W_{\sR}(C)=\text{Tr}_{\sR}\left({\cal P} \exp\le(- \oint_{C} A\ri){\cal P} \exp\le(-\oint_{C} \bar A\ri)  \right)~,
\end{equation}
and $C$ is a closed loop in the bulk of AdS$_3$. This observable is fully gauge-invariant, and will typically be an interesting observable if the bulk loop wraps some non-trivial object in the bulk (e.g. the horizon of a BTZ black hole). Note that the trace involves a sum over the infinitely many states of the highest-weight representation. 
 
We may also consider an open-ended Wilson {\it line} operator. To define this object we specify the locations of its endpoints $(x_i, x_f)$. We must also specify boundary data in the form of two specific states $|U_i\rangle, |U_f\rangle \in \sR$ at these endpoints. We may then define the following operator:
\be\label{eq:wilson1}
W_{\sR}(x_i,x_f)= \langle U_f | {\cal P} \exp\le(- \int_{\gamma} A\ri){\cal P} \exp\le(-\int_{\gamma} \bar A\ri) | U_i \rangle ~,
\ee
where now $\gamma(s)$ is a curve with bulk endpoints $(x_i,x_f)$ parametrized by $s$. 
$W_{\sR}(x_i, x_f)$ is no longer fully gauge-invariant; clearly it depends in a gauge-covariant manner on the choice of boundary data $|U_i\rangle$, $|U_f\rangle$. Nevertheless, for flat connections, $W_{\sR}(x_i,x_f)$ only depends on the topology of $\gamma$, but not on the shape of the curve.


From a geometric point of view, the Wilson line described above describes the physics of a massive point particle propagating from $x_i$ to $x_f$ on AdS$_3$. A point particle in the classical limit is characterized by at least one continuous parameter: the mass $m$. This data is stored in the choice of highest-weight representation $\sR$ that defines the Wilson line. Further details of this representation are given in full detail in Sec. \ref{sec:hilbert}. For now we require only that the representation is specified by two constants $(h,\bar h)$ which determine the Casimirs of the $sl(2)$ algebra.  Their identification with the mass $m$ and orbital spin $\hat s$ of the particle is given by
\be\label{eq:cms}
  m^2= c_2 + \bar c_2~,\quad \hat s=\bar h-h~,
\ee
where $c_2=2h(h-1)$ and $\bar c_2=2\bar h(\bar h-1)$ are the quadratic Casimirs; note that the AdS radius is set to unity.

From the point of view of AdS/CFT, the developments in \cite{Ammon:2013hba,deBoer:2013vca,Castro:2014mza,deBoer:2014sna,Hegde2016} show that if the endpoints $x_i$, $x_f$ are taken to infinity, the  Wilson line operator defined in \eqref{eq:wilson1} is a bulk observable that computes correlation functions of light operators $\langle \Psi| \sO (y_i) \sO(y_f)|\Psi\rangle $ in the dual CFT. Here $|\Psi \rangle$ is a ``heavy'' state whose gravitational dual is given by the bulk connections $(A,\bar A)$ and $\sO (y)$ is a ``light'' operator whose scaling dimensions $(h,\bar h)$ are encoded in the choice of representation\footnote{ Here light denotes an operator that, as the central charge $c$ goes to infinity, its conformal weight is fixed,  while a heavy operator has a scaling dimension that is linear with $c$. Equivalently, in gravity we would say that it is a particle with a small mass in Planck units.}  
 $\sR$. 
 
 In what follows we  limit the discussion to $h=\bar h$; see \cite{Castro:2014tta,Castro2016} for a discussion when $\hat s\neq 0$. 


\subsection{Path integral representation of the Wilson line}\label{sec:pi}

This particular Wilson line is somewhat more complex than those normally studied in compact gauge theories, simply due to the fact that $\sR$ has infinitely many states in it. We now review the work of \cite{Ammon:2013hba}, who 
constructed $\cal R$ as the Hilbert space of an auxiliary quantum mechanical system that lives on the Wilson line, replacing the trace over $\sR$ by a path integral over a worldline field $U$. We pick the dynamics of $U$ so that upon quantization the Hilbert space of the system is the desired representation $\CR$.
More concretely, we rewrite \eqref{Wilsonfunc} as
\begin{equation}\label{pathint}
 W_{\sR}(x_i,x_f)=\int\mathcal{D}U \, e^{-S(U,A,\bar{A})_{\gamma}}~, 
\end{equation}
where the the auxiliary system can be described by the following action:
\bea\label{actionfin}
S(U, A,\bar{A})_{\gamma}&=&\sqrt{c_2} \int_\gamma ds \sqrt{\Tr \le(U^{-1}{D_sU}\ri)^2} 
\eea
The variable $s$ parametrizes the curve ${\gamma}$, and we pick $s\in [s_i,s_f]$. Here the trace $\text{Tr}(...)$ is a short-cut notation for the contraction using the Killing forms, i.e. if $P \in sl(2,\mathbb{R})$
\be
\text{Tr}(P^2)=\eta_{ab}P^{a}P^{b}~,
\ee
where $P=P^aL_a$ and $L_a$ is a generator of $sl(2,\mathbb{R})$. There is also a (classically) equivalent first-order formulation of this action that is more convenient for certain applications (such as the generalization to higher spin gravity). In the first order formulation it is manifest that $c_2$ is the Casimir of the representation, and satisfies $c_2 = 2h(h-1)$.  This action requires that $h=\bar h$. As the entire action is multiplied by a factor of $\sqrt{c_2}$, $h \to \infty$ defines a semi-classical limit of the path integral, and for the remainder of this section we will follow \cite{Ammon:2013hba} and work only in this limit. In subsequent sections we relax this restriction.   

This action is invariant under a local $\slt \times \slt$ symmetry: in particular the covariant derivative is defined as
\be
D_sU\equiv \frac{d}{ds}U+A_sU-U\bar{A}_s\,, \quad A_s\equiv A_{\mu}\frac{dx^{\mu}}{ds}~,\quad \bar A_s\equiv \bar A_{\mu}\frac{dx^{\mu}}{ds}~, \label{gaugecov} 
\ee
where $A(x)$ and $\bar{A}(x)$ are the connections that determine the background, and in the action \eqref{actionfin} they are pulled back to the worldline $x^{\mu}(s)$. Under an $\slt \times \slt$ gauge transformation by finite group elements $L(x)$, $R(x)$, the gauge fields transform as
\bea\label{gaugeA}
A_{\mu}(x)&\rightarrow& L(x)(A_{\mu}(x)+\partial_{\mu})L^{-1}(x)\,,\cr \bar{A}_{\mu}(x)&\rightarrow& R^{-1}(x)(\bar{A}_{\mu}(x)+\partial_{\mu})R(x) \ . 
\eea
The worldline action is then invariant under the following transformation of the worldline field:
\bea\label{symloc}
U(s)&\rightarrow& L(x^\mu(s))U(s)R(x^\mu(s))
\eea

%
Now for an open ended Wilson line as in \eqref{eq:wilson1}, we must still specify boundary data on $U(s)$ at the endpoints of the curve.\footnote{For a closed Wilson loop as in \eqref{Wilsonfunc}, we simply require that the field $U(s)$ be single-valued.} We thus pick two $\slt$ elements $U_i, U_f$ and require that $U(s = s_i) = U_i$, $U(s = s_f) = U_f$. For a semi-classical level this is sufficient, and in later sections we will explain in detail the relationship between this choice of boundary data and the quantum states $|U_i\rangle$ and $|U_f\rangle$ defined in \eqref{eq:wilson1}.  

We now consider the evaluation of this Wilson line on a fixed classical background defined by $A$ and $\bar{A}$. In the $h \to \infty$ limit, this can be done by evaluating the on-shell action \eqref{actionfin} for the field $U(s)$ subject to the boundary conditions described above. This computation was explained in detail in \cite{Ammon:2013hba}. Here we write the answer in a way that will generalize simply to our results in the next section. In particular the answer only depends on the $\slt$ evolution of the state from the starting point to the endpoint. If we thus consider flat connections
  \be\label{gs11}
 A(x) = g_L(x) d g_L(x)^{-1}\,, \quad  \bar A(x) = g_R(x)^{-1} d g_R(x)\,, 
 \ee
and the following group elements
\be
g_L(x_f) g_L(x_i)^{-1}= \mathcal{P} \exp\le(-\int_{x_i}^{x_f} A\ri)~, \qquad g^{-1}_R(x_f) g_R(x_i)= \mathcal{P} \exp\le(-\int_{x_i}^{x_f} \bar{A}\ri)~,
\ee
then the on-shell action $S$ can be written as
\be
S_{\mbox{on-shell}} = \sqrt{\frac{c_2}{2}}\al~, \qquad  V \exp(-\al L_0) V^{-1} \equiv  g_L(x_f) g_L(x_i)^{-1} U_i  g_R(x_i)^{-1}  g_R(x_f)U_f^{-1}~,  \label{aldef} 
\ee
 where  $\al$ labels the conjugacy class of the group element $g_L(x_f) g_L(x_i)^{-1} U_i  g_R(x_i)^{-1}  g_R(x_f)U_f^{-1}$.
The Wilson line \eqref{pathint} in this state is then given by
\begin{equation}\label{pathint}
W_{\sR}(x_i,x_f)  =\exp\le(-\sqrt{\frac{c_2}{2}}\al\ri)~.
\end{equation}
Note that the role of the boundary data $U_i, U_f$ in \eqref{aldef} is to tie together the two sectors, left and right; we will return to this point in what follows. 

\subsection{Geometric interpretation: proper distances} \label{sec:geod}


So far, our review has been very abstract, with no physical interpretation given to $A$ and $\bar{A}$. However we know that for appropriate choices of these gauge connections, this system should represent the physics of a particle moving on AdS$_3$; we now explain how the result above is related to geometry. In particular, $\al$ defined in \eqref{aldef} turns out to be related to the proper distance from $x_i$ to $x_f$.

To understand this, note that the action \eqref{actionfin} can be suggestively written as
\be
S = \sqrt{c_2}\int_\gamma ds \sqrt{\Tr \le(\le(A_{\mu} - \tilde{\bar{A}}_\mu\ri)\le(A_{\nu} - \tilde{\bar{A}}_{\nu}\ri)\frac{dx^{\mu}}{ds} \frac{dx^{\nu}}{ds} \ri)},
\ee
where the dependence on $U(s)$ in \eqref{eomfull} has been hidden in the definition of $\tilde{{\bar A}}_{\nu}$:
\be
\tilde{\bar{A}}_s \equiv  U\bar A_s U^{-1}- \frac{d}{ds} U \, U^{-1}  ~.
\ee
Note that if we now define a {\it generalized} vielbein\footnote{Technically we can only define the components of the vielbein along the trajectory; in the considerations of this section this does not matter.} along the trajectory as 
\be
e_{\mu} = \frac{1}{2} \le(A_{\mu} - \tilde{\bar{A}}_{\mu}\ri)
\ee
then we may write the action very simply in terms of the metric associated to this vielbein as $g_{\mu\nu} = 2 \;\Tr_f e_{\mu} e_{\nu}$, i.e. 
\bea
S = {\sqrt{2c_2}} \int_{\gamma}ds \sqrt{g_{\mu\nu}(x) \frac{dx^{\mu}}{ds}\frac{dx^{\nu}}{ds}}~, \label{distance}
\eea
which is manifestly the proper distance associated to the metric $g_{\mu\nu}$. Thus the Wilson line is probing a geometry that is assembled in a particular manner from the connections $A, \bar{A}$, where the dynamics of the auxiliary field $U$ is playing a role in tying together the two connections into a vielbein. Note that the prefactor $\sqrt{c_2}$ indicates that the value of the Casimir controls the bulk mass of the probe, as we alluded to previously. 

We also consider the equations of motion obtained from varying \eqref{actionfin} with respect to $U$:
\be
\frac{d}{ds}\le((A - \tilde{\bar{A}})_{\mu} \frac{dx^{\mu}}{ds}\ri) + [\tilde{\bar{A}}_{\mu}, A_{\nu}]\frac{dx^{\mu}}{ds}\frac{dx^{\nu}}{ds} = 0\label{eomfull} \ .
\ee
Normally one considers these as equations for $U(s)$: nevertheless, if one fixes $U(s)$ and thinks of the variable as being the choice of {\it path} $x^{\mu}(s)$, then this is precisely the geodesic equation for the metric $g_{\mu\nu}$. 
%
From here it is clear that the value of the Wilson line between any two bulk points is
\be\label{eq:wsaddle}
W_{\sR}(x_i, x_f) \sim \exp\le(-2h D(x_i, x_f)\ri),
\ee
where $D(x_i, x_f)=2\al$ is the length of the bulk geodesic connecting these two points. Here `$\sim$' denotes the limit of large $c_2$, where $c_2=2h(h-1)\sim 2h^2$,  and hence the classical saddle point approximation is valid. 

In what follows we will provide a proper quantum-mechanical treatment of this Wilson line. 

\section{Hilbert space representation}   \label{sec:hilbert}

The path integral approach to evaluate \eqref{eq:wilson1} provides insight into the transformation properties for the field $U$: this choice is in great part responsible of the geometric interpretation of $W_{\sR}(x_i,x_f)$  in AdS$_3$ gravity. Based on this, in this section we will carefully explain the relationship between the field $U(x)$ and quantum mechanical states in the highest-weight representation. This will allow us to evaluate $W_{\sR}(x_i,x_f)$ without the need of taking a classical limit --in contrast to \eqref{eq:wsaddle}-- and, in later sections, have a refined geometric and holographic interpretation of our Wilson line. 

\subsection{Highest weight representations}


We first review some facts associated with highest-weight representations. Some words on notation are appropriate: when we are discussing an abstract realization of the $\aslt$ algebra with no particular representation in mind, we will denote the generators with capital $L_a$. We denote the generators of $\aslt$ acting on the highest weight state by $\ell_a$. A highest-weight representation is defined with respect to a reference state $|h\rangle$ that is an eigenstate of $\ell_0$ and is annihilated by $\ell_1$:
\be
\ell_0 |h\rangle = h |h \rangle~, \qquad \ell_{1} |h\rangle = 0~.
\ee
We may now define excited states by acting on $|h\rangle$ with $\ell_{-1}$, and the correctly normalized states are defined by
\begin{align}
\ell_{-1} |h,k\rangle  = \sqrt{(k+1)(k+2h)} |h, k+1\rangle~, \quad\qquad \ell_1 |h,k\rangle  = \sqrt{k(k+2h-1)}|h,k-1\rangle~, \label{norm1}
\end{align}
where the state $|h,k\rangle$ has $L_0$ eigenvalue $(k+h)$: i.e. $k$ counts the energy {\it above} the ground state, and $|h,0\rangle = |h\rangle$. The Casimir of this representation is $2h(h-1)$:
\be
\eta^{ab} \ell_a \ell_b |h,k\rangle = 2h(h-1)|h, k\rangle~, \label{casimir} 
\ee
where $\eta^{ab}$ is the Killing form. 

We will be interested in states that transform in a highest-weight representation under a tensor product of two independent copies of $\aslt \times \aslt$ with $h=\bar h$, and so we will label them as $|h,k\rangle\otimes |h,\bar k\rangle \equiv |h;k,\bar k\rangle$, where the ground state is $|h,0,0\rangle$. We denote the $\aslt$ generators acting on the first $k$ index (the ``left'') by $\ell_a$ and those acting on the $\bar k$ index (the ``right'') as $\bell_a$. These form two independent $\aslt$ algebras, and we have
\be
[\ell_a, \bell_a] = 0~.
\ee


The group action of each copy  of $\slt$ on these states is the usual one: in particular, we have
\be
G(M^{-1})\ell_a G(M) = D_{a}^{\phantom{a}a'}(M) \ell_{a'}~, \qquad \bG(M^{-1}) \bell_a \bG(M) = D_{a}^{\phantom{m}a'}(M)\bell_{a'}~, \label{sltac}
\ee  
with the $D$'s the representation matrices for the adjoint representation of $\aslt$. Note that we have
\be
G(M_1)G(M_2) = G(M_1 M_2)~, \qquad D_{a}^{\phantom{m}b}(M_1) D_{b}^{\phantom{m}c}(M_2) = D_{a}^{\phantom{m}c}(M_1 M_2)~.
\ee



\subsection{Rotated Ishibashi states}\label{sec:rotatedishi}

We will now define a family of quantum states that have the same transformation as the classical field $U(x)$ in \eqref{symloc}.
To do so it is convenient to consider the following triplet of $\aslt$ operators, labeled by an element $U \in \slt$:
\be\label{eq:defQa}
Q_a(U) \equiv \ell_{a} + D_{a}^{\phantom{m}a'}(U)\bell_{a'}~.
\ee
This is a linear combination of the generators on the two sides, with one side rotated by $U$. We will denote a state that is annihilated by $Q_a(U)$ for all $a$ as $|U\rangle$, i.e.
\be
Q_a(U) |U\rangle = 0 ~.\label{Vdef} 
\ee
This defines a \emph{rotated state}, each labeled by an element $U$ of $\slt$. We now explore some of the properties of these states. First consider commuting $G(L)\bG(R)$ through $Q_m(U)$. We find
\be\label{eomstate}
G(L)\bG(R^{-1}) Q_a(U) =  D_{a}^{\phantom{m}a'}(L^{-1}) Q_{a'}(L U R) G(L)\bG(R^{-1})~.
\ee
Acting with this relation on the state $|U\rangle$, we find that the state $G(L) \bG(R^{-1}) | U \rangle$ is annihilated by $Q_{a}(L U R)$. But by the definition of the $U$ states, this means that 
\be
G(L) \bG(R^{-1}) |U\rangle = |L U R \rangle~. \label{Utrans}
\ee
Thus we see that acting on a $U$ state with an element of $\slt \times \slt$ causes it to transform inhomogenously precisely as the {\it classical} $U$ field did in \eqref{symloc}. We also note that every $U$ state is left invariant under {\it some} diagonal subgroup of $\slt \times \slt$, that with $L = U R^{-1} U^{-1}$.  


It will be useful to have some explicit examples of $|U\rangle$  in terms of the highest weight representation discussed above. As a start let us consider the state $|U\rangle=|\Sigma_{\rm Ish}\rangle$ whose action on the generators is
\be\label{eq:darot}
D_{a}^{\phantom{m}a'}(\Sigma_{\rm Ish})\,\bell_{a'}= \Sigma_{\rm Ish}\, \bell_{a} \,\Sigma_{\rm Ish}^{-1}=-\bell_{-a}~,
\ee
and as a group element is
\be
\Sigma_{\rm Ish} \equiv \exp\le(-i\frac{\pi}{2}(L_{1}-L_{-1})\ri)~. \label{SigdefIsh} 
\ee
 Using \eqref{eq:darot}, \eqref{Vdef} becomes
\be\label{Ishieq}
\le(\ell_{a} - \bell_{-a}\ri)|\Sigma_{\rm Ish}\rangle = 0~.
\ee
This equation has the following unique solution, 
\be\label{Ishi}
|\Sigma_{\rm Ish}\rangle = \sum_{k=0}^\infty  |h;k,k\rangle~,
\ee
which is (in its Virasoso incarnation \cite{Ishibashi:1988kg}) called the ``Ishibashi state.'' 
Another choice for our states is setting $|U\rangle=|\Sigma_{\rm cross}\rangle$ whose action is
\be\label{eq:darot1}
D_{a}^{\phantom{m}a'}(\Sigma_{\rm cross})\,\bell_{a'}= \Sigma_{\rm cross}\, \bell_{a}\, \Sigma_{\rm cross}^{-1}=-(-1)^a\bell_{-a}~,
\ee
and as a group element it reads
\be
\Sigma_{\rm cross} \equiv \exp\le(\frac{\pi}{2}(L_1 + L_{-1})\ri)~. \label{Sigdefcross} 
\ee
For this choice \eqref{Vdef} becomes
\be\label{Ishieq1}
\le(\ell_{a} - (-1)^a \bell_{-a}\ri)|\Sigma_{\rm cross}\rangle = 0~,
\ee
and the unique solution to this equation is
\be\label{cross}
|\Sigma_{\rm cross}\rangle = \sum_{k=0}^\infty (-1)^k |h;k,k\rangle~,
\ee
which is usually referred to as the ``crosscap (or twisted) Ishibashi state'' \cite{Ishibashi:1988kg}.  The state $|\Sigma_{\rm cross}\rangle$ (rather than $|\Sigma_{\rm Ish}\rangle $) will play an important role in section \ref{sec:examples}, for reasons that we will elaborate on there.

 If we can construct any reference state in this family, then we can find any other state by acting on it with an appropriately chosen $G(L)$ and/or $\bar G(R^{-1})$.\footnote{Note that we are allowed to rotate a $|U\rangle$ state if $G(L)$ has a well defined action on the representation. This implies that not any rotation is allowed. For example, we cannot rotate $|\Sigma_{\rm cross}\rangle$ to the state $|U=\mathds{1}\rangle$, which is ill defined since setting $U= \mathds{1}$ in \eqref{Vdef} has no solution in the highest weight representation. The reason is that $\Sigma_{\rm cross}$ is an outer automorphism: it has a well defined action on the group elements as signalled by \eqref{eq:darot1}, but not on the states of representation (it would flip the sign of $L_0$ eigenvalue). Similar statements hold for $\Sigma_{\rm Ish}$.}  And for this reason we will call the states $|U\rangle$ (in a slight abuse of notation) rotated Ishibashi states. Our rotated Ishibashi states are coherent states that live in the product of two highest weight representations and only involve the global part of the conformal group, unlike the states used for boundary CFT \cite{Ishibashi:1988kg,Cardy:2004hm}.



\subsection{Inner product}\label{sec:innerprod}

An important object in our analysis is the inner product of a rotated Ishibashi state. These states are not orthogonal-- they form an overcomplete basis-- which leads to a non-trivial expression. The relevant matrix element to evaluate any such inner product is
\be\label{eq:matrixelement}
 \langle \Sigma | G(L) \bG(R^{-1}) |\Sigma \rangle~,
\ee
where  $ |\Sigma \rangle$ is a reference state from our family of rotated Ishibashi states. For concreteness we will   take  $ |\Sigma \rangle$ to be either
\be
  |\Sigma_{\rm Ish} \rangle ~~{\rm or}~~  |\Sigma_{\rm cross} \rangle ~,
\ee
as defined in \eqref{Ishi} and \eqref{cross}. 

Evaluating \eqref{eq:matrixelement} leads to
\begin{align}\label{eq:steps}
\langle \Sigma | G(L) \bG(R^{-1}) |\Sigma \rangle&= \langle \Sigma | G(L \,\Sigma \,R\, \Sigma^{-1}) |\Sigma \rangle\cr
 &=\sum_{k=0}^{\infty}  |a_k|^2 \langle h, k  | G(L \,\Sigma \,R\,\Sigma^{-1}) |h, k\rangle\cr
 &=\sum_{k=0}^{\infty} \langle h, k  | G(L \,\Sigma \,R\, \Sigma^{-1}) |h, k\rangle\cr
 &=\sum_{k=0}^{\infty} \exp(-\al (k+h)) = \frac{e^{-\al h}}{1-e^{-\al}} ~.
\end{align}
In the first equality we used \eqref{Utrans}. In the second line we used  \eqref{Ishi} and \eqref{cross}; the coefficient $a_k$ is equal to 1 and $(-1)^k$, respectively. In the third line we used that $|a_k|^2=1$, which reduces the computation to a trace of the group element inside the bracket. In the last line we decomposed the group element as
\be\label{innerproduct0}
L \,\Sigma \,R\, \Sigma^{-1} \equiv V \exp(-\al L_0) V^{-1} ~,
\ee
where $\al$ controls the conjugacy class of the group element in question. The last equality is our final result, which is just a $\aslt$ character of $ G(L \,\Sigma \,R\, \Sigma^{-1}) $. From here the role of $ |\Sigma \rangle$ is becoming more evident: it controls how the right element $R$ would act as left element relative to $L$ and vice versa. 

The result \eqref{eq:steps} immediately generalizes to the inner product between any two of the $U$-states as defined in \eqref{Vdef}: any rotated state continuously connected to $\Sigma$ will satisfy
\be
\langle U_1 | U_2 \rangle =  \frac{e^{-\al h}}{1-e^{-\al}}~, \qquad U_1^{-1} U_2 \equiv V \exp(-\al L_0) V^{-1}~. \label{genans} 
\ee
In other words, the inner product between any two $U$-states $U_1$ and $U_2$ is a function only the ``magnitude'' $\al$ of the conjugacy class of the group element that relates $U_1$ to $U_2$. $\al$ can be thought of as an invariant distance between the two elements on the group manifold (and indeed we will develop its geometric interpretation in the next subsection). Note that as $U_1$ approaches $U_2$, $\al \to 0$ and thus the norm of any $U$ state itself is infinite: this divergence can be seen immediately from noting that the norm of $ |\Sigma \rangle$ diverges. 

Finally, the $U$ states satisfy a completeness relation. It is shown in Appendix \ref{app:completeness} through explicit computation that for $2h > 1$ we have
\be
\int dU  |U \rangle \langle U | = \frac{(2\pi)^2}{2(2h-1)} \mathds{1}~, \label{completeness} 
\ee
where $dU$ is the Haar measure on $\slt$. In pedestrian terms, this simply means that we treat $\slt$ as being locally AdS$_3$ and integrate over it using the usual volume measure, taking care to integrate over $\slt$ and not over its universal cover.

\subsection{The Green's function on the group manifold} \label{sec:waveq}
Here we discuss a few further properties of the inner product $\langle U_1 | U_2 \rangle$ computed above. In particular, the inner product \eqref{genans} is actually a Green's function with respect to the invariant Laplacian on the $\slt$ group manifold. 

We begin by placing coordinates $\sig^{\al}$ on the group manifold $\slt$. Let us denote the usual generators of $sl(2,\mathbb{R})$ in the fundamental representation by $L_a$. As $\slt$ is a group manifold, there exist vector fields $\xi^{\al}_{a}$ and $\bar{\xi}^{\al}_a$ that generate the group action on a point in the manifold from the left and from the right, i.e.
\be
\xi^{\al}_a \frac{\p U(\sig)}{\p \sig^{\al}} = L_a U(\sig)~, \qquad \bar{\xi}^{\al}_a \frac{\p U(\sig)}{\p \sig^{\al}} = U(\sig) L_a~. \label{killingfield}
\ee
As the $U$-states \eqref{Utrans} transform in the same way, they satisfy:
\be
\xi^{\al}_a \p_{\al} |U(\sig) \rangle  = | L_a U(\sig) \rangle = \ell_a |U(\sig)\rangle ~,
\ee
as well as a similar relation for the barred sector. Now we act with this relation twice on the $\sig_2$ coordinates parametrizing the inner product $\langle U(\sig_1) | U(\sig_2) \rangle$ with $U(\sig_1) \neq U(\sig_2)$. In particular, denote the Killing form on $sl(2,\mathbb{R})$ by $\eta^{ab}$ and compute
\be
\eta^{ab} \langle U(\sig_1) | \xi^{\al}_a \p_{\al} \le(\xi^{\beta}_b \p_{\beta} |U(\sig_2)\ri) \rangle = \eta^{ab} \langle U(\sig_1) | \ell_b \ell_a | U(\sig_2)\rangle = 2h(h-1) \langle U(\sig_1) | U(\sig_2) \rangle~, \label{waveq}
\ee 
where in the last equality we have used the Casimir relation \eqref{casimir}. It is straightforward to verify that the second-order differential operator on the left-hand side of \eqref{waveq} is (up to a factor of $\ha$) the invariant Laplacian on $\slt$, which we denote by $\Box_U$. As our analysis holds only for non-coincident $U_1$, $U_2$, we conclude that
\be
\le(\ha \Box_{U_2} - 2h(h-1)\ri)\langle U_1 | U_2 \rangle = \frac{1}{8\pi} \delta(U_1, U_2)~. \label{groupwaveq}
\ee
Here $\delta(U_1,U_2)$ is a delta function on the group manifold that is nonzero only if $U_1 = U_2$, and which is normalized to satisfy $\int dU \delta(U_0, U) = 1$ with $dU$ the Haar measure on $\slt$ and $U_0$ a reference group element. This can of course also be checked by explicitly verifying that \eqref{genans} satisfies the appropriate Laplacian; this is also the fastest way to verify the existence of the delta function on the right-hand side.

\subsection{Relationship to path integral}
In this section we will demonstrate that the in the large-$h$ limit, the inner product defined above can be computed from a path integral over a classical field $U(s)$, as used in \cite{Ammon:2013hba} and reviewed in Sec. \ref{sec:path}. Essentially we will perform the analogue of the usual construction of the path integral for quantum mechanical systems, where the non-compact nature of the representation, and therefore of the $U$ states, provide some extra wrinkles. 

Consider computing an inner product of the form
\be
\langle U_f | G(L) \bar{G}(R^{-1}) | U_i \rangle~. \label{inprod}
\ee
To give this a quantum-mechanical interpretation, we will represent the group elements $L$ and $R$ as path-ordered exponentials of gauge fields $A(s)$ and $\bA(s)$, where $s$ should be thought of as ``time'', i.e. 
\be
L = \mathcal{P} \exp\le(-\int_{s_i}^{s_f} A_s ds\ri)~, \qquad R^{-1} = \mathcal{P} \exp\le(-\int_{s_i}^{s_f} \bA_s ds\ri)~. 
\ee
To make contact with conventional quantum mechanics, one can imagine that $A$ and $\bA$ define a Hamiltonian for the system defining time-evolution along $s$. We will now derive a path integral expression for the inner product \eqref{inprod}. We follow the normal algorithm of dividing the path from $s_i$ to $s_f$ into many small intervals of size $\ep$, discretizing the path as $s_{i}, s_{i+1}, s_{i+2} \cdots s_{f-1}, s_{f}$, where the time step is $s_{j} - s_{j-1} = \ep$.  

We may then break up each path-ordered exponential:
\be
\mathcal{P} \exp\le(-\int_{s_i}^{s_f} A_{s} ds\ri) = e^{-\ep  A_{s}(s_f)} e^{-\ep  A_{s}(s_{f-1})} \cdots e^{-\ep  A_{s}(s_i)} = \prod_{j} e^{-\ep  A_{s}(s_j)}~,
\ee 
and similarly for the right sector. The inner product takes the form
\be
\langle U_f | G(L) \bar{G}(R^{-1}) | U_i \rangle = \langle U_f |\prod_{j}\le[G\le(e^{-\ep  A_{s}(s_j)}\ri)\bG\le(e^{-\ep \bA_{s}(s_j)}\ri)\ri] | U_i \rangle~.
\ee
We now use \eqref{completeness} to insert a complete set of $U$ states at each time step. We find
\be
\langle U_f | G(L) \bar{G}(R^{-1}) | U_i \rangle = \mathcal{N} \langle U_f |\prod_{j}\le[G\le(e^{-\ep  A_{s}(s_j)}\ri)\bG\le(e^{-\ep \bA_{s}(s_j)}\ri) \int dU |U(s_j)\rangle \langle U(s_j)| \ri]| U_i \rangle~,
\ee
where we have introduced an overall prefactor $\mathcal{N}$ to absorb factors of the form $(2h-1)^{\infty}$ into the usual ambiguities in the measure of the path integral. We see that we must evaluate many inner products of the form
\be
\langle U(s_{j+1}) | G\le(e^{-\ep  A_{s}(s_j)}\ri)\bG\le(e^{-\ep \bA_{s}(s_j)}\ri) | U(s_j) \rangle~.
\ee
To evaluate this inner product, we make the usual assumption that most contributions to the path integral come from reasonably smoothly varying $U(s)$, so that we may assume that $U(s_{j+1}) = U(s_j) + \ep \frac{dU(s_j)}{ds} + \sO(\ep^2)$. Thus to lowest order in $\ep$ we are evaluating
\be
\bigg\langle U(s_{j})\le(\mathds{1} + U(s_{j})^{-1} \ep \frac{dU(s_{j})}{ds}\ri)\bigg|G\le(e^{-\ep  A_{s}(s_j)}\ri)\bG\le(e^{-\ep \bA_{s}(s_j)}\ri)\bigg|U(s_j) \bigg\rangle~. 
\ee
We use the transformation property of the $U$ states \eqref{Utrans} to move all of the group elements to the ket on the right to obtain
\be
\bigg\langle U(s_j) \bigg|e^{-\ep A_{s}(s_j)} U(s_j) e^{-\ep \bA_s(s_j)}\le(\mathds{1}- U(s_j)^{-1}\ep \frac{dU}{ds}(s_j)\ri)\bigg\rangle \ . 
\ee
Next, we use the general form for the inner product \eqref{genans} to conclude that 
\be
\langle U(s_{j+1}) | G\le(e^{-\ep  A_{s}(s_j)}\ri)\bG\le(e^{-\ep \bA_{s}(s_j)}\ri) | U(s_j) \rangle = \frac{e^{-\al(s_j) h}}{1-e^{-\al(s_j)}}~,
\ee
where $\al(s_j)$ is given by the conjugacy class of the $\slt$ element
\be\label{eq:msj}
M(s_j) \equiv \exp\le(-\ep\le(U^{-1}\frac{dU}{ds} + U^{-1}A_s U - \bA_s\ri)\ri)\bigg|_{s = s_j} \qquad M(s_j) = V^{-1}\exp(-\al(s_j) L_0) V~,
\ee
where to obtain this expression we expanded all terms up to order $\ep$, and then re-exponentiated the resulting expression. It should be understood that this expression is correct only up to order $\ep$. We have encountered a version of \eqref{eq:msj} in \eqref{innerproduct0} and \eqref{genans}, and we  will encounter again in subsequent sections. The a simple way to read off  $\al(s_j)$ is by noticing that \eqref{eq:msj} --and its counsins \eqref{innerproduct0} and \eqref{genans}-- are independent of the $sl(2)$ representation. With these freedom, we choose to solve this equation in the fundamental representation of $sl(2)$, described by the $2\times2$ traceless matrices, where $\al$ is given by a trace:
\be
\al(s_j) = 2 \ep \sqrt{\Tr_f\le(U^{-1} {D_sU}\ri)^2}\bigg|_{s = s_j} \label{alphabit}~.
\ee
Here the gauge-covariant derivative ${D_sU}$ is that defined in \eqref{gaugecov}, and our conventions for the fundamental representation are given in appendix \ref{app:sl2}. 

We have thus computed the contribution of one infinitesimal piece of the path. Assembling all of these pieces by taking the product, we see that the full inner product \eqref{inprod} is given by
\be
\langle U_f | G(L) \bar{G}(R^{-1}) | U_i \rangle = \mathcal{N}\prod_j \le(\int dU(s_j) \frac{e^{-\al(s_j) h}}{1-e^{-\al(s_j)}}\ri)\bigg|_{U(s_i) = U_i, U(s_f) = U_f}~.
\ee
We now consider taking the continuum limit $\ep \to 0$; the product of integrals $dU(s_j)$ over each group element at each point on the path becomes a path integral $[\mathcal{D} U]$ over a continuous worldline field $U(s)$. 
We first consider the numerator of the above expression: this naturally becomes an integral over a smooth action: 
\be
\prod_j \exp\le(- h \al(s_j)\ri) \to \exp\le(-2 h \int_{s_i}^{s_f} ds \sqrt{\Tr_f\le(U^{-1} {D_sU}\ri)^2}\ri)~,
\ee
i.e. precisely the exponential of the action $S[U]$ postulated on physics grounds in \cite{Ammon:2013hba}. 

We now turn to the denominator $1-e^{-\al(s_j)}$. In the limit $\ep \to 0$, each $\al(s_j)$ is infinitesimal, and thus we may write: 
\be
\prod_j (1-e^{-\al(s_j)})^{-1} \approx \prod_j (\al(s_j))^{-1} = \prod_j \sqrt{\frac{\ep}{2\pi}} \int d\sig(s_j) \exp\le(-\frac{\ep}{2} \sig(s_j)^2 \al(s_j)^2\ri)~,
\ee
where we have introduced a new auxiliary field $\sig(s_j)$ at each point on the worldline; integrating out this field generates the denominator (up to an overall ill-defined prefactor that depends on the discretization). The full path integral is thus
\be
\int_{U(s_i) = U_i}^{U(s_f) = U_f} [\mathcal{D}U] \exp\le(-S[U,\sig]\ri)~. \label{pathint}
\ee
where the full continuum action is
\be
S[U,\sig] =  \int_{s_i}^{s_f} ds \le(2 h \sqrt{\Tr_f\le(U^{-1} {D_sU} \ri)^2} + \ha \sig(s)^2\;\Tr_f\le(U^{-1} {D_sU}\ri)^2\ri)~.
\ee

In the $h \to \infty$ limit, we may ignore the second term in the action: this is then precisely the path integral \eqref{pathint}-\eqref{actionfin} which was proposed on symmetry grounds in \cite{Ammon:2013hba}.  

We can now see that at finite $h$, the path integral proposed in \cite{Ammon:2013hba} must be corrected by additional ``quantum'' terms arising from the measure of the path integral when integrating over $U$ states. This additional term --the wrinkle we alluded to at the start of this subsection-- arises from the fact that the inner product of two nearby $U$ states is divergent, which is itself a direct consequence of the non-compactness of $\slt$ and the resulting infinite tower of highest weight states. It would be interesting to understand better the physical significance of this term; however in this paper we will not attempt to treat the path integral \eqref{pathint} at finite $h$, and will instead simply directly compute matrix elements from the algebraic approach developed above.

\section{Wilson lines: Local Fields and Geometry}\label{sec:examples}

Our goal in this section is to give a geometric interpretation to the algebraic construction  in  Sec. \ref{sec:hilbert}.  We will start in Sec. \ref{sec:gi}, by going through the simple exercise of casting our gravitational Wilson line in \eqref{eq:wilson1} along the lines of  the discussion in Sec. \ref{sec:innerprod}. In Sec. \ref{sec:ph} we will argue that for invertible connections $(A,\bar A)$, we can interpret the transformation properties of the group elements in the Wilson line as moving the endpoints of the operator in AdS$_3$. This justifies the geometric interpretation of the algebraic object. And finally, in Sec. \ref{sec:btzads} we will show how to build a local bulk field from our rotated Ishibashi states; these constructions will be explicitly done for global AdS and the static BTZ black hole.

\subsection{Gravitational Wilson line as an overlap of two states}\label{sec:gi}

The results in Sec. \ref{sec:hilbert} gives a prescription to evaluate overlap of states in the highest weight representation. In this section we would like to implement those results to a gravitational Wilson line. More concretely, we would like to analyse 
\be\label{Iwilsonmatch}
W_{\mathcal{R}}(x_f, x_i) = \langle  \Sigma | \,G\le(\sP e^{-\int_{x_i}^{x_f} A}\ri)\,\,\bar G\le(\, \sP e^{-\int_{x_i}^{x_f} \bA}\ri) | \Sigma \rangle \,,
\ee
 as an overlap of a suitable initial and final $|U\rangle$ state. We keep the reference state $|\Sigma\rangle$ generic so far, and we will discuss the different choices  $\Sigma_{\text{Ish}}$, and $\Sigma_{\text{cross}}$ in Sec. \ref{sec:btzads}. As in Sec. \ref{sec:path}, $\gamma(s)$ is a curve with bulk endpoints $(x_i,x_f)$; we use the affine parameter $s \in [s_i,s_f]$ where $x(s = s_i) = x_i$ and $x(s = s_f) = x_f$.

 To recast \eqref{Iwilsonmatch} as an inner product,  it is useful to rewrite the flat connections as
  \be\label{gs}
 A(x) = g_L(x) d g_L(x)^{-1}\,, \quad  \bar A(x) = g_R(x)^{-1} d g_R(x)\,, 
 \ee
 Using the transformation of the path ordered exponential under \eqref{gs}:
  \be\label{eq:aba1}
  \mathcal{P}\,e^{-\int_\gamma {A}}= g_L(x_f) g_L(x_i)^{-1}\,, \qquad  \mathcal{P}\,e^{-\int_\gamma {\bar A}}= g^{-1}_R(x_f) g_R(x_i)~,
  \ee
and therefore
\be\label{Iwilsonmatch2}
\langle \Sigma |G\left(\mathcal{P}\,e^{-\int_\gamma {A}}\right) \bar G\left( \mathcal{P}\,e^{-\int_\gamma {\bar A}}\right) |\Sigma\rangle=\langle \Sigma |G\left(g_L(x_f) g_L(x_i)^{-1}\right) \bar G\left( g^{-1}_R(x_f) g_R(x_i)\right)  |\Sigma\rangle\,,
\ee
 To write this expression as an overlap between to states, we  define
\be\label{Ustate}
 |U(x)\rangle\equiv G\left(g_L(x)^{-1}\right) \bar G\left( g_R (x)\right)|\Sigma\rangle\,.
  \ee
%
%
%
and with this, we can rewrite the previous amplitude as 
\be\label{Iwilsonmatch3}
\langle \Sigma |G\left(\mathcal{P}\,e^{-\int_\gamma {A}}\right) \bar G\left(\,\mathcal{P}\,e^{-\int_\gamma {\bar A}}\,\right)| \Sigma\rangle=\langle U(x_f)  |U(x_i)\rangle\,.
\ee
It is important to note that in this expression we have implicitly assumed that the group element $g_L$ obeys 
$$g_L^{-1}= g_L^\dagger~,$$
and similarly for $g_R$. All of our manipulations will use group elements that are unitary. And we should stress that $|U(x)\rangle$ is not gauge invariant. In its definition in \eqref{Ustate} we implicitly made a choice: we are splitting the path from $x_i$ to $x_f$ to a mid point where $g_L=g_R = \mathds{1}$, and without any further specification of the connections, we have not motivated nor justified this choice.  This bug does not affect \eqref{Iwilsonmatch3}, and we will ignore it for now.  We will return to this point in Sec. \ref{sec:btzads} when we directly analyse $|U(x)\rangle$.


Having casted the gravitational Wilson line as an inner product in \eqref{Iwilsonmatch3}, we can now use the same logic that leads to   \eqref{eq:steps} and \eqref{innerproduct0}. In particular we find that 
\be\label{finalWilson}
\langle \Sigma |G\left(\mathcal{P}\,e^{-\int_\gamma {A}}\right) \bar G\left(\,\mathcal{P}\,e^{-\int_\gamma {\bar A}}\,\right)| \Sigma\rangle = \frac{e^{-\al(x_i,x_f) h}}{1-e^{-\al(x_i,x_f)}}\,
\ee
where, following \eqref{innerproduct0} for this case, $\alpha(x_i,x_f)$ is given by the solution to
\be\label{finalWilson2}
g_L(x_f) g_L(x_i)^{-1} \tilde g_R(x_i)^{-1} \tilde g_R(x_f) = V \exp(-\al(x_i,x_f) L_0) V^{-1}\,.
\ee
and we define
\be
\tilde g_R\equiv \Sigma^{-1} g_R \Sigma~, \qquad \tilde\bA\equiv \Sigma^{-1}\bar A \Sigma ~.
\ee
Note that while, by definition, $A$ and $\bA$ act on different spaces, the role of $\Sigma$ is to tie together these two sectors; $\tilde\bA$ can be thought of as the `left' version of the `right' connection.

To solve for $\alpha(x_i,x_f)$ in \eqref{finalWilson2}, it is useful to note that this equation is independent of the $sl(2,\R)$ representation, and hence we can simply use a finite dimensional representation.\footnote{There is an ambiguity in the sign in front of $\alpha$ when \eqref{finalWilson2} is considered in a finite dimensional representation. However, $\alpha>0$ as required by the convergence of  \eqref{eq:steps}.} Using the fundamental representation of $sl(2,\mathbb{R})$ (see appendix \ref{app:sl2}), and after taking the trace both sides of \eqref{finalWilson2}, gives
\be\label{alpha}
\cosh\le(\frac{\alpha(x_i,x_f)}{2}\ri)=\left(\frac{1}{2}\text{Tr}_f\left(g_L(x_f) g_L(x_i)^{-1} \tilde g_R(x_i)^{-1} \tilde g_R(x_f)\right)\right)\,.
\ee
where $\text{Tr}_f$ is the trace in the fundamental representation. Using \eqref{gs} together with \eqref{glgr} and \eqref{sl2conn2}, we find that $\alpha(s_i,s_f)=2 D(s_i,s_f)$ is the geodesic length of an effective metric given by  
 \be\label{eq:effmetric}
 g_{\mu\nu}={1\over 2} {\rm Tr}(A_\mu -\tilde \bA_\mu )(A_\nu -\tilde \bA_\nu)~.
 \ee
The relevant metric for global AdS and BTZ is given in \eqref{Banadosmet}. Therefore, the inner product is
\begin{align}\label{SCHlength}
\langle \Sigma |G\left(\mathcal{P}\,e^{-\int_\gamma {A}}\right) \bar G\left(\,\mathcal{P}\,e^{-\int_\gamma {\bar A}}\,\right)| \Sigma\rangle = \frac{e^{-2hD(x_i,x_f) }}{1-e^{-2D(x_i,x_f)}}\,.
\end{align}
This is the familiar bulk-to-bulk propagator of a minimally coupled scalar field in a locally AdS$_3$ background \cite{Berenstein:1998ij,Danielsson:1998wt}.  In the semi-classical limit, where the numerator is negligible and $h$ is large, the saddle point approximation of the path integral in \eqref{SCHlength} precisely agrees with \eqref{eq:wsaddle}.  The background metric \eqref{eq:effmetric} is in agreement with \eqref{distance}, and \eqref{finalWilson2} is equivalent to \eqref{aldef}.

At the level of evaluating \eqref{SCHlength}, the detailed nature of $| \Sigma\rangle$ can be overlooked: provided the endpoint states satisfies
\be\label{eq:rot1}
G(L)G(R^{-1})| \Sigma\rangle =| L\Sigma R\rangle~,
\ee
 we will obtain \eqref{SCHlength}, and interpret it as the bulk-to-bulk propagator of a scalar field with background metric \eqref{eq:effmetric}. With this perspective, if the input is $g_{\mu\nu}$, we could just infer the values of $(A,\tilde\bA)$ and use them in \eqref{SCHlength}, without making explicit reference to the difference between $\bA$ and $\tilde\bA$, and hence neglect the role of $| \Sigma\rangle$. However, $|U(x)\rangle$ is an object sensitive to $|\Sigma\rangle$, and as we will discuss  in section \ref{sec:btzads},  this will disentangle the different features that  $|\Sigma\rangle$ captures as we build local probes in AdS$_3$.

 \subsection{Algebra meets geometry}\label{sec:ph}
 
An expression such as \eqref{SCHlength} makes rather evident that the Wilson line is a propagator, and hence its ties to geometry. The drawback however is the brut aspect of the observation: it relied on evaluating explicitly the observable on AdS$_3$ and the BTZ background. In this section we will do better. We will show that the object
\be
W_{\mathcal{R}}(x_f, x_i) = \langle \Sigma | G\le(\sP e^{-\int_{x_i}^{x_f} A}\ri) \bG\le(\sP e^{-\int_{x_i}^{x_f} \bA}\ri) | \Sigma \rangle \label{bulkdef}
\ee
can be understood as a bulk-to-bulk propagator with respect to the bulk spacetime metric associated with the flat connections $A, \bA$. 
The important improvement here relative to our prior observations is that here we treat the Wilson line quantum mechanically, and as such it will capture the geometry as perceived by a {\it bulk field} of an arbitrary mass. 

We begin by assuming that the bulk spacetime is simply connected (e.g. for pure AdS$_3$). In this case all paths from $x_i$ to $x_f$ are topologically equivalent, and \eqref{bulkdef} is a well-defined function of the two endpoints. 

We first recall that in \eqref{groupwaveq} it was shown that the object $\langle U_1 | U_2 \rangle$ was a Green's function on the group manifold $\slt$. This is logically distinct from showing that the matrix element \eqref{bulkdef} is a Green's function on the bulk metric defined by $A, \bA$. 

To make a connection between these two objects, we first need to establish how the matrix elements in \eqref{bulkdef} change if we move, for instance, the point $x_f$. The  dependence on endpoints $x_i$ and $x_f$ enters in \eqref{bulkdef} as follows: using \eqref{gs}-\eqref{eq:aba1}, the matrix element reads
\begin{align}\label{eq:wr2}
W_{\mathcal{R}}(x_f, x_i) &= \langle  \Sigma | G(g_L(x_f) g_L(x_i)^{-1}) \bar G( g_R(x_f)^{-1} g_R(x_i)) | \Sigma\rangle\cr
&= \langle \Sigma| G(g_L(x_f) g_L(x_i)^{-1}  \tilde g_R(x_i)^{-1} \tilde g_R(x_f)) | \Sigma \rangle~.
\end{align}
In the second line we made use of the transformation properties of our reference states \eqref{Utrans}, and used the definition $\tilde g_R\equiv \Sigma^{-1} g_R \Sigma$ . We note that this is where the choice of $|U\rangle$ to be rotated states is crucial: the state combines both sectors, which will lead to a geometric interpretation of $W_{\mathcal{R}}(x_f, x_i)$ in the subsequent steps. From \eqref{eq:wr2}, the full dependence on $x_i$ and $x_f$ enters through the following group element
\be
\sG(x_f,x_i) \equiv g_L(x_f) g_L(x_i)^{-1} \tilde g_R(x_i)^{-1} \tilde  g_R(x_f)~.
\ee
 Taking an $x_f$ derivative of this group element, we have
\be
\frac{\p}{\p x_f^{\mu}} \sG(x_f,x_i) = -A_{\mu}(x_f) \sG(x_i, x_f) + \sG(x_i, x_f) \tilde \bA_{\mu}(x_f)~. \label{movement} 
\ee
Recall now that \eqref{groupwaveq} was shown by exploiting the fact that the left and right action of the group generated a set of vector fields on the group manifold \eqref{killingfield}. We would now like to extend this idea to the geometric bulk, i.e. we seek a set of vector fields $\zeta^{\mu}_a, \bar{\zeta}^{\mu}_a$ defined on AdS$_3$ such that
\be
\zeta^{\mu}_a \frac{\p}{\p x_f^\mu} \sG(x_f, x_i) = L_a \sG(x_f, x_i)~, \qquad \bar{\zeta}^{\mu}_a \frac{\p}{\p x_f^{\mu}} \sG(x_f, x_i) = \sG(x_f, x_i)L_a~. \label{killingbulk} 
\ee
Multiplying both sides of these equations by $L_b$ and taking a trace, we see that the defining relations become
\begin{equation} 
\zeta^{\mu}_a \;\tr_f\le(\le(-A_{\mu} + \sG \tilde\bA_{\mu} \sG^{-1}\ri)L_b\ri) = \eta_{ab}~, \qquad
\bar{\zeta}^{\mu}_a \;\tr_f\le(\le(-\sG^{-1} A_{\mu} \sG + \tilde\bA_{\mu} \ri)L_b\ri) = \eta_{ab} ~.\ee
These equations will have solutions for $\zeta$, $\bar{\zeta}$ if the $3 \times 3$ matrices (with rows labaled by $\mu$ and columns by $b$) multiplying them from the right are invertible. However from \eqref{vbspin}, we see that these matrices are closely related to the usual vielbein $e \sim A - \tilde \bA$ in the metric formulation of 3d gravity, with one side rotated by the $\slt$ transformation defined by $\sG(x_f, x_i)$. The condition that the generalized vielbeins above be invertible appears to be required for a simple geometric interpretation of the bulk spacetime.  

If the generalized vielbeins shown above are invertible, then the $\zeta, \bar{\zeta}$ exist, and we have shown that movement in bulk spacetime is equivalent to movement on the group manifold. Furthermore the condition \eqref{killingbulk} guarantees that they satisfy the $sl(2,\mathbb{R}) \times sl(2,\mathbb{R})$ algebra as Killing vectors on the bulk spacetime. Thus following through the same steps as in \eqref{groupwaveq}, we conclude that
\be
\le(\ha\Box_{x_f} - 2h(h-1) \ri)W_{\mathcal{R}}(x_f, x_i) = \frac{1}{8\pi} \frac{\delta(x_f, x_i)}{\sqrt{-g}}
\ee
where now $\Box_{x_f}$ is the Laplacian on the bulk AdS$_3$ spacetime.  The construction of $\zeta, \bar{\zeta}$ will be carried out explicitly in Sec. \ref{sec:btzads}.

We now consider the case where the bulk spacetime is not simply connected, e.g. the BTZ black hole. For a black hole the bulk connections have a nontrivial holonomy around the black hole horizon. In this case the definition of the open-ended Wilson line $W_{\mathcal{R}}(x_f, x_i)$ in \eqref{bulkdef} is incomplete: as there are multiple inequivalent bulk paths that connect $x_i$ and $x_f$, we must specify a path, and different choices of path will result in different answers. 

In this case, if we would like to obtain an unambiguous answer that depends only on the endpoints, one prescription is to sum over all inequivalent paths, i.e., we define a path-summed Wilson line as 
\be
{W}_{\mathcal{R}}(x_f,x_i) = \sum_{C(x_f,x_i)} W_{\mathcal{R}}(C(x_f,x_i))
\ee
where the sum is over all topologically inequivalent paths $C(x_f, x_i)$ that connect $x_f$ to $x_i$. An example of such situation is nicely capture by the BTZ black hole. In this case the inequivalent paths correspond to geodesics winding around the horizon multiple times, and the resulting propagator is a sum over these windings. For the static black hole, the resulting propagator is
\be
{W}_{\mathcal{R}}(x_f,x_i)_{\rm BTZ} = \sum_{n\in \ZZ}\frac{e^{-2hD_n(x_i,x_f) }}{1-e^{-2D_n(x_i,x_f)}}~,
\ee
with
\be
D_n(x_i,x_f)=\frac{1}{r_+^2}\left( r_fr_i\cosh (r_+\Delta\phi+ 2\pi r_+ n )- \sqrt{(r_f^2-r_+^2)(r_i^2-r_+^2)}\cosh (r_+\Delta t)\right)~.
\ee
Here we are using the geodesic length in \eqref{geodlengthr}, and $n$ controls the number of times the path encloses the horizon. In the metric formulation this sum can be understood as the sum over images that gives the propagator the correct periodicity condition (see e.g. \cite{Maldacena:2001kr}), which in complete agreement with our expression.

\subsection{Local fields}\label{sec:btzads}

 In the last portion of this section we will evaluate and interpret $| U(x)\rangle$ as defined in \eqref{Ustate}. As mentioned there, this definition is gauge dependent. A definition of $| U(x)\rangle$ that reinstates this dependence is
 \begin{align} \label{Ustate22} 
 | U(x)\rangle= G\left(g_L(x_0)g_L(x)^{-1}\right) \bar G\left( g^{-1}_R(x_0) g_R (x) \right)|\Sigma\rangle~.
 \end{align}
where $x_0^\mu$ is a bulk reference point where $ | U(x_0)\rangle=|\Sigma\rangle$. In other words, the point $x_0^\mu$ defines where in the bulk we should locate the state $|\Sigma\rangle$. Once this choice is made, $| U(x)\rangle$ is a  prescription on how to move through the bulk the state $|\Sigma\rangle$ from $x_0^\mu$ to a point $x^\mu$.

We will decompose the state \eqref{Ustate22} as a sum over local functions in the infinite-dimensional representation
\be\label{changeb}
 | U(x)\rangle=\sum_{k,\bar k=0}^{\infty} \Phi^*_{k,\bar k}(x)|h,k,\bar k \rangle~,
\ee
and evaluate $ \Phi_{k,\bar k}(x)$. Alternatively, the function $\Phi_{k,\bar k}(x)$ is
\be\label{innerbasis}
\Phi_{k,\bar k}(x)=\langle U(x)| h,k,\bar k \rangle=\langle h,k,\bar k| U(x)\rangle^{\dagger}\,.
\ee
This function is the object that will provide local bulk information in the Chern-Simons formulation of 3D gravity.

The explicit calculation of this  $\Phi_{k,\bar k}(x)$  can be a complicated task. A way to proceed is by using the technique in Appendix A of \cite{Miyaji:2015fia}. The aim there is to find a differential operators $\mathcal{L}_{a}(x)$ whose action in the inner product \eqref{innerbasis} is 
\be\label{L1}
\langle U(x) |\ell_{a}| h,k,\bar k \rangle=\mathcal{L}_{a}(x)\,\langle U(x) | h,k,\bar k \rangle \,.
\ee
where $\ell_{a}$ is the infinite-dimensional generator that acts as in \eqref{norm1}, and $\mathcal{L}_{a}(x)$ is a differential operator acting on the $x$ variables, whose explicit form depends on the state $| U(x)\rangle$. Analogous formulas can be found for the barred sector. These operators are precisely the vector fields introduced in \eqref{killingbulk}, i.e. we have
\be
{\cal L}_a(x) = \zeta^{\mu}_a \frac{\p}{\p x^\mu}~, \qquad  \bar{\cal L}_a(x) =\bar\zeta^{\mu}_a \frac{\p}{\p x^\mu}~.
\ee
Equation \eqref{L1}, together with \eqref{norm1}, implies that
\begin{align}\label{raiseAdSfield}
\mathcal{L}_{-1}\Phi_{k,\bar k}(x)= \sqrt{(k+1)(k+2h)} \Phi_{k+1,\bar k}(x)\,,\cr  \mathcal{\bar L}_{-1}  \Phi_{k,\bar k}(x)= \sqrt{(\bar k+1)(\bar k+2h)}  \Phi_{k,\bar k+1}(x)\,. 
\end{align}
 $\Phi_{0, 0}(x)$ can be fully determined by solving following differential equations
\be\label{hwcond}
\mathcal{L}_{0}(x) \Phi_{0, 0}(x)= h\Phi_{0,0}(x)\,,\qquad \mathcal{L}_{1}(x) \Phi_{0, 0}(x) =0\,,
\ee
together with its barred version. Therefore, we will be able to infer the form of $\Phi_{k,\bar k}(x)$, by successively applying $\mathcal{L}_{-1}(x)$, and $\mathcal{\bar L}_{-1}(x)$ to the seed $\Phi_{0, 0}(x)$.  From here it follows that $\Phi_{k,\bar k}(x)$ obeys the Casimir equation
\be\label{casimir}
\left(\mathcal{L}^2(x)+ \mathcal{\bar L}^2(x)\right)\Phi_{k,\bar k}(x)= 4h(h-1)\Phi_{k,\bar k}(x)\,,
\ee
where $\mathcal{L}^2=-(\mathcal{L}_{-1}\mathcal{L}_{1}+\mathcal{L}_{1}\mathcal{L}_{-1})+2\mathcal{L}_0^2$.  In other words, $\Phi_{k,\bar k}(x)$ is a local bulk field of mass $m^2=4h(h-1)$ and whose boundary conditions are given by the highest weight conditions \eqref{raiseAdSfield}-\eqref{hwcond}.

Finally, once we have the explicit expression of the functions $\Phi_{k,\bar k}(x)$, we will compute the inner products of two states \eqref{changeb} as
\be\label{innergen}
\langle  U(x_f)  | U(x_i)\rangle=\sum_{k,\bar k}^{\infty} \Phi_{k,\bar k}(x_f)  \Phi^*_{k,\bar k}(x_i)\,.
 \ee
Note that when we evaluate \eqref{innergen} we will not make use of \eqref{eq:steps}, and hence the derivations in this portion give an alternative and more direct derivation of \eqref{SCHlength}. In the following, we will carry out this procedure for two explicit backgrounds. Sec.  \ref{sec:global} is devoted to global AdS$_3$, which agrees completely with the results in\cite{Miyaji:2015fia}, and Sec. \ref{sec:BTZ} focuses on the static BTZ black hole.

\subsubsection{Global $AdS_3$}\label{sec:global}

Let us consider the state $|U\rangle$  for global AdS$_3$ and build explicitly $ \Phi_{k,\bar k}(x)$ for this background.  To start we will first infer the group elements from the standard metric for AdS$_3$, i.e.
\be\label{eq:AdSgl-1}
ds^2=-\cosh^2\rho \,dt^2+{d\rho^2}+\sinh^2\rho \,d\phi^2\,.
\ee
Using \eqref{eq:effmetric},  it is straight forward to read from \eqref{eq:AdSgl-1} unitary group elements $g_L$ and $\tilde g_R$. Details are presented in App. \ref{sec:coord}, and the resulting elements are
\be\label{eq:ads22}
g_L(x)= e^{(\ell_{1}-\ell_{-1})\rho/2} e^{-i\ell_0x^{+}} ~,\qquad \tilde g_R(x)= e^{-i\ell_0x^{-}}e^{(\ell_{1}-\ell_{-1})\rho/2}~, 
\ee
where $x^\pm=t\pm \phi$.
We will use the definition \eqref{Ustate22} with $g_L(x_0)=\mathds{1}=g_R(x_0)$; this places $|\Sigma\rangle$ at the origin of AdS in accordance with the results in \cite{Verlinde:2015qfa,Miyaji:2015fia,NakayamaOoguri2015}. This gives 
\begin{align}\label{Ustateglobal}
 | U(x)\rangle_{\rm AdS}&= G\left(g_L(x)^{-1}\right) \bar G\left( g_R (x) \right)|\Sigma\rangle\cr
&= G\left(g_L(x)^{-1} \tilde g_R (x)^{-1} \right) |\Sigma\rangle \cr
&=e^{ix^+ \ell_0} e^{-{\rho} (\ell_1-\ell_{-1} )} e^{ix^- \ell_0} |\Sigma\rangle ~,
\end{align}
where we used \eqref{Utrans} and \eqref{eq:ads22}. For most of the following derivations we will drop the subscript ``AdS'' and restore it when needed.

The next step is to find the differential operators $\mathcal{L}_a(x)$ in  \eqref{L1} for global AdS$_3$. For that we use the inner product as
\be\label{innerbasisgl}
\Phi_{k,\bar k}(x)=\langle U(x)| h,k,\bar k \rangle=\langle \Sigma |e^{-ix^- \ell_0}e^{{\rho} (\ell_1-\ell_{-1} )} e^{-ix^+ \ell_0}|h,k,\bar k \rangle\,,
\ee
%
Taking derivatives with respect to the global coordinates gives
  \begin{align}\label{gendif0}
\partial_{x^{+}}\langle U(x) | h,k,\bar k \rangle  &= -i\langle \Sigma |e^{-ix^- \ell_0}e^{{\rho} (\ell_1-\ell_{-1} )} e^{-ix^+ \ell_0}\ell_0|h,k,\bar k \rangle\, \, ,\nonumber\\
\partial_{\rho}\langle U(x) | h,k,\bar k \rangle  &= \langle \Sigma |e^{-ix^- \ell_0}e^{{\rho} (\ell_1-\ell_{-1} )} (\ell_1-\ell_{-1} ) e^{-ix^+ \ell_0}|h,k,\bar k \rangle\, \, ,\nonumber\\
\partial_{x^{-}}\langle U(x) | h,k,\bar k \rangle  &= -i\langle \Sigma |e^{-ix^- \ell_0}\ell_0e^{{\rho} (\ell_1-\ell_{-1} )} e^{-ix^+ \ell_0}|h,k,\bar k \rangle\, ,
\end{align}
and using commutation relations, we can move the generators that are not in the exponents to the right, to get
  \begin{align}\label{gendif0b}
  \partial_{x^{+}}\langle U(x) | h,k,\bar k \rangle  &= -i\langle U(x)|\ell_0|h,k,\bar k \rangle\, \, ,\nonumber\\
\partial_{\rho}\langle U(x) | h,k,\bar k \rangle  &=\langle U(x)|(e^{-ix^+}\ell_1-e^{ix^+}\ell_{-1} )|h,k,\bar k \rangle\, \, ,\nonumber\\
\partial_{x^{-}}\langle U(x) | h,k,\bar k \rangle  &= -i\langle U(x)| \cosh 2\rho \,\ell_0+ \frac{\sinh 2\rho}{2}(e^{-ix^+}\ell_1+e^{ix^+}\ell_{-1} )|h,k,\bar k \rangle\, .
\end{align}
Now, it is straight forward to obtain the differential operators that follow \eqref{L1} for global AdS$_3$; these read 
  \begin{align}\label{gendif}
\mathcal{L}_{0}  &= i \partial_{x^+} \, ,\nonumber\\
\mathcal{L}_{\pm1} &=i e^{ \pm ix^{+}} \left[  \frac{\cosh 2\rho}{\sinh 2\rho}\partial_{x^{+}} -\frac{1}{\sinh 2\rho}\partial_{x^{-}}\mp\frac{i}{2}\partial_{\rho} \right]\,.
\end{align}
It is important to remark that these differential operators were built without making direct reference to $|\Sigma\rangle$. 

To find the barred differential operators we follow a procedure analogous to what we did in \eqref{gendif0}-\eqref{gendif0b}, but using the following inner product:
\be\label{innerbasisglb}
\Phi_{k,\bar k}(x)=\langle U(x)| h,k,\bar k \rangle=\langle \Sigma |\Sigma^{-1}e^{ix^+ \bell_0}e^{-{\rho} (\bell_1-\bell_{-1} )} e^{ix^- \bell_0}\Sigma|h,k,\bar k \rangle\,,
\ee
where we are rewriting the action of the left group elements as an action via the right, i.e.
\begin{align}\label{Ustateglobal1}
 | U(x)\rangle_{\rm AdS}&= G\left(g_L(x)^{-1}\right) \bar G\left( g_R (x) \right)|\Sigma\rangle\cr
&=\bar G\left( \Sigma^{-1} \tilde g_R (x)  g_L(x)\Sigma \right) |\Sigma\rangle ~.
\end{align}
While in \eqref{innerbasisgl} we could ignore $\Sigma$, we are now forced to understand how $\Sigma$ acts on the states to infer the  differential operators $\mathcal{\bar L}_{a}$. A sensible choice is to require that  $\mathcal{\bar L}_{a}$ are related to  $\mathcal{L}_{a}$ by replacing ${x^+}\leftrightarrow {x^-}$, i.e.
  \begin{align}\label{gendifb}
\mathcal{\bar L}_{0}  &= i \partial_{x^-} \,,\\
\mathcal{\bar L}_{\pm1} &=i e^{ \pm ix^{-}} \left[  \frac{\cosh 2\rho}{\sinh 2\rho}\partial_{x^{-}} -\frac{1}{\sinh 2\rho}\partial_{x^{+}}\mp\frac{i}{2}\partial_{\rho} \right] \,.
\nonumber
\end{align}
This is the familiar assignment of Killing vectors in AdS$_3$; the interesting twist here is that not any choice of probe $\Sigma$ will achieve this assignment.
A choice of $|\Sigma\rangle$ that delivers  \eqref{gendifb} for the group element \eqref{innerbasisglb} is the crosscap state in \eqref{cross}: 
\be\label{choice:cross}
|\Sigma\rangle=|\Sigma_{\text{cross}}\rangle~.
\ee 
The Ishibashi state $|\Sigma_{\text{Ish}}\rangle$ has a different effect. Using  \eqref{eq:darot}  in \eqref{innerbasisglb} would lead to operators $\mathcal{\bar L}_{a}$ that are related to  $\mathcal{L}_{a}$ through ${x^{+}}\leftrightarrow {x^{-}}$, and $\rho\rightarrow -\rho$. Still, the resulting differential operators are Killing vectors in the metric \eqref{eq:AdSgl-1}, and they follow the $sl(2,\RR)_L\times sl(2,\RR)_R$ algebra. Therefore, the choice of the Ishibashi state has also an interpretation in the geometrical description of fields in AdS$_3$. The only minor drawback is that  the relation of Killing vectors and algebra elements has a different normalization relative to the standard choices in global AdS$_3$.

Our starting point in this subsection was the metric for AdS$_3$ in \eqref{eq:AdSgl-1}. Another starting point is to use the fact that global AdS$_3$ is maximally symmetric, and the group elements that label rotations and translations in this space are
\begin{align}\label{Ustateglobale}
 | U(x)\rangle_{\rm AdS}=e^{i\ell_0x^{+}}\,e^{i\bell_0x^{-}}e^{-\frac{\rho}{2} (\ell_1-\ell_{-1} +\bell_1-\bell_{-1} )}  | \Sigma_{\text{cross}} \rangle\,,
\end{align}
as it was done in \cite{Miyaji:2015fia,Goto:2017olq}. For the crosscap state, \eqref{Ustateglobale} is in complete agreement with \eqref{eq:ads22}. The choice $| \Sigma_{\text{Ish}} \rangle$  would lead to different group elements, which is tied to the fact that in this case we have a non-stantard relation between algebra elements and Killing vectors of the geometry.


  The differential operators \eqref{gendif}, and  \eqref{gendifb} are Killing vectors of global AdS$_3$, as advocated in Sec. \ref{sec:ph}. Moreover,  $\left(\mathcal{L}^2(x)+ \mathcal{\bar L}^2(x)\right)$ in \eqref{casimir} is the usual d'Alembertian for AdS$_3$. Therefore, $\Phi_{k,\bar k}(s)$ is a scalar field with mass $m^2=4h(h-1)$ in a global AdS background.  Now, we can solve \eqref{hwcond} using the previous differential operators, as done in \cite{Maldacena:1998bw}; the highest weight state is
\begin{align}\label{hwglobal}
 \Phi_{0, 0}(x)=\langle U(x) | h,0,0 \rangle=\frac{ e^{-2 i h t }}{(\cosh\rho)^{2h}}\,.
\end{align}
To find $\Phi_{k,\bar k}(x)$ we simply need to identify the solutions to \eqref{casimir} and organize them as $\mathcal{L}_{-1}(x)$, and $\mathcal{\bar L}_{-1}(x)$ acting on  \eqref{hwglobal}. This leads to
\be\label{AdSfield}
\Phi_{k,\bar k}(x)= C_{k,\bar k}\, e^{-ih(x^++x^-)}e^{-i(k x^++\bar k x^-)}(\tanh \rho)^{\bar k- k}(\cosh \rho)^{-2h}P_k^{(\bar k-k,\, 2h -1)}(1-2\tanh^2\rho)\,,
\ee
 where $P_n^{(a,\, b)}$ are Jacobi polynomials, and $C_{k,\bar k}=(-1)^k\sqrt{\frac{k! (2 h+\bar k-1)!}{\bar k! (2 h+k-1)!}}$ is a constant that has been chosen to match the normalizations in \eqref{raiseAdSfield}. Therefore, we found the state \eqref{changeb} in a global AdS background.  This is in complete agreement with the known results of normalizable wavefunction in AdS$_3$ as in, e.g., \cite{Balasubramanian:1998sn}.

 We are ready to compute the overlap of two states at different positions in the bulk. Using \eqref{innergen} with \eqref{AdSfield} gives
\begin{align}\label{sumint}
\langle  U(x_f)  | U(x_i)\rangle
=&\sum_{k,\bar k=0}^{\infty}e^{-ih(\Delta x^++\Delta x^-)}e^{-i( k \Delta x^++ \bar k \Delta x^-)}{\frac{k! (2 h+\bar k-1)!}{\bar k! (2 h+k-1)!}}(\tanh{\rho_f}\tanh{\rho_i})^{\bar k- k}(\cosh{\rho_f}\cosh{\rho_i})^{-2h}
\nonumber\\
&\times P_{k}^{(\bar k- k,\, 2h -1)}(1-2\tanh^2\rho_f)P_{ k}^{(\bar k- k,\, 2h -1)}(1-2\tanh^2\rho_i)\,.
\end{align}
The previous sum is performed in the Appendix \ref{sec:sum}. If we choose $x=\tanh^2\rho_i$, $y=\tanh^2\rho_f$, $r=e^{-i\Delta x^-}$, and $s=e^{-i\Delta x^+}$, the left hand side of \eqref{mastersum2} is equal to \eqref{sumint}. Applying \eqref{mastersum2}, we find 
\be\label{resultglobal}
\langle  U(x_f)  | U(x_i)\rangle= \frac{\left(\sigma(x_i,x_f)+\sqrt{\sigma^2(x_i,x_f)-1}\right)^{-(2h-1)}}{2\sqrt{\sigma^2(x_i,x_f)-1}}=\frac{e^{-2hD(x_i,x_f) }}{1-e^{-2D(x_i,x_f)}}\,.
\ee
where $D(x_i,x_f)$ is the geodesic length of global AdS, given in \eqref{geodlength} with $C=-1/4$. This in complete agreement with the result in \eqref{SCHlength}.

\subsubsection{BTZ}\label{sec:BTZ}

As we did for global AdS$_3$, we will now find the local functions $\Phi_{k,\bar k}(x)$ for the static BTZ background. Our starting point is to build the group elements $(g_L,\tilde g_R)$ from the metric, which for the black hole reads
\be\label{BTZmet1}
ds^2=-(r^2-r_+^2)dt^2+\frac{dr^2}{r^2-r_+^2}+r^2d\phi^2\,.
\ee
In Appendix \ref{sec:coord} we build the appropriate connections for the black hole are \eqref{sl2conn2} that are compatible with \eqref{BTZmet1} and unitary in the highest weight representation. The resulting BTZ state is\footnote{Following the discussion around \eqref{Ustate22} and \eqref{eq:ads22}, we have chosen here $g_L(x_0)=\mathds{1}=\tilde g_R(x_0)$. In contrast to global AdS$_3$, there is no physical motivation to make this choice for BTZ: it simply makes some of the subsequent manipulations easier. It would be interesting to investigate what is a physically sound choice of $x_0^\mu$ in future work. }
\begin{align}\label{UstateBTZ}
| U(x)\rangle_{\rm BTZ}&=
G\left(g_L(x)^{-1} \tilde g_R (x)^{-1} \right)|\Sigma\rangle \nonumber\\
&=e^{-\frac{i}{4}\left((8C-2)\ell_0-(4C+1)(\ell_{1}+\ell_{-1})\right)x^{+}}e^{-{\rho} (\ell_1-\ell_{-1})} e^{-\frac{i}{4}\left((8C-2)\ell_0+(4C+1)(\ell_{1}+\ell_{-1})\right)x^{-}} |\Sigma\rangle~,
\end{align}
where we casted all the elements as acting on the left, and we introduced
\be
r=r_+\cosh^2(\rho-\rho_*)~,\qquad 4C= e^{2\rho_*}= r_+^2~, \qquad x^\pm=t\pm\phi~.
\ee
%
Following the same procedure as in Sec. \ref{sec:global}, we can find differential operators defined as \eqref{L1} for the BTZ state. Using\footnote{For simplicity, we will omit the subscript `BTZ' in most of this section, and restore it when needed. } 
\begin{align}\label{innerbasisBTZ}
\Phi_{k,\bar k}(x)&=\langle U(x)| h,k,\bar k \rangle\\
&=\langle \Sigma |e^{\frac{i}{4}\left((8C-2)\ell_0+(4C+1)(\ell_{1}+\ell_{-1})\right)x^{-}}e^{{\rho} (\ell_1-\ell_{-1})}e^{\frac{i}{4}\left((8C-2)\ell_0-(4C+1)(\ell_{1}+\ell_{-1})\right)x^{+}}|h,k,\bar k \rangle\,,\nonumber
\end{align}
%
we find the non-barred differential operators 
 \begin{align}\label{gendifBTZ}
\mathcal{L}_{0}  &= - 2\alpha_+ \sqrt{C} \sinh \left(2 \sqrt{C}x^+\right)\partial_{\rho} +\left(\alpha_-+\frac{1+f(\rho )^2}{2f(\rho )} \alpha_+ \cosh \left(2 \sqrt{C}x^+\right)\right)\partial_{x^+} \nonumber\\\nonumber
&+\frac{1-f(\rho )^2}{2f(\rho )} \alpha_+ \cosh \left(2 \sqrt{C}x^+\right)\partial_{x^-}~,\\\nonumber
\mathcal{L}_{\pm1} &=\left( \pm\frac{1}{2} \cosh \left(2 \sqrt{C}x^+\right)-2 \alpha_- \sqrt{C} \sinh \left(2 \sqrt{C}x^+\right)\right)\partial_{\rho}\\
&+\left(\alpha_++\frac{1+f(\rho )^2}{8 \sqrt{C}f(\rho )} \left(\alpha_- 4 \sqrt{C}\cosh \left(2 \sqrt{C}x^+\right)\mp\sinh \left(2 \sqrt{C}x^+\right)\right)\right)\partial_{x^+}\nonumber\\
&+\left(\alpha_++\frac{1-f(\rho )^2}{8\sqrt{C}f(\rho )}\left(\alpha_-4 \sqrt{C} \cosh \left(2 \sqrt{C}x^+\right)\mp\sinh \left(2 \sqrt{C}x^+\right)\right)\right)\partial_{x^-}\, .
\end{align}
with
\be
f(\rho)\equiv\frac{e^{2 \rho }-4 C}{4 C+e^{2 \rho }}\,,\qquad \alpha_{\pm}\equiv\frac{i (4 C\pm 1)}{16 C}\,.
\ee
In order to obtain the barred generators, we proceed as done for global AdS$_3$ in \eqref{innerbasisglb}-\eqref{Ustateglobal1}, i.e. we rewrite the state $|U(x)\rangle_{\rm BTZ}$ as having an action only via right group elements. This gives
\begin{align}\label{innerbasisBTZb}
\Phi_{k,\bar k}(x)&=\langle U(x)| h,k,\bar k \rangle\\
&=\langle \Sigma |\Sigma^{-1} e^{\frac{i}{4}\left(-(8C-2)\bell_0+(4C+1)(\bell_{1}+\bell_{-1})\right)x^{+}}e^{-{\rho} (\bell_1-\bell_{-1})}e^{\frac{i}{4}\left(-(8C-2)\bell_0-(4C+1)(\bell_{1}+\bell_{-1})\right)x^{-}}\Sigma|h,k,\bar k \rangle\,.\nonumber
\end{align}
As before, we will fix $\Sigma$  such that the barred differential operators, $ \mathcal{\bar L}_{a} $, are equal to the non-barred operators with ${x^+}\leftrightarrow {x^-}$, as it is natural in the metric formulation. A quick inspection singles out $ | \Sigma_{\text{cross}}\rangle$ as the appropriate choice rather than $ | \Sigma_{\text{Ish}}\rangle$. Using $| \Sigma\rangle=| \Sigma_{\text{cross}}\rangle$ in \eqref{innerbasisBTZb} we find
 \begin{align}\label{gendifBTZa}
\mathcal{\bar L}_{0}  &= - 2\alpha_+ \sqrt{C} \sinh \left(2 \sqrt{C}x^-\right)\partial_{\rho} +\left(\alpha_-+\frac{1+f(\rho )^2}{2f(\rho )} \alpha_+ \cosh \left(2 \sqrt{C}x^-\right)\right)\partial_{x^-} \nonumber\\\nonumber
&+\frac{1-f(\rho )^2}{2f(\rho )} \alpha_+ \cosh \left(2 \sqrt{C}x^-\right)\partial_{x^+}~,\\\nonumber
\mathcal{\bar L}_{\pm1} &=\left( \pm\frac{1}{2} \cosh \left(2 \sqrt{C}x^-\right)-2 \alpha_- \sqrt{C} \sinh \left(2 \sqrt{C}x^-\right)\right)\partial_{\rho}\\
&+\left(\alpha_++\frac{1+f(\rho )^2}{8 \sqrt{C}f(\rho )} \left(\alpha_- 4 \sqrt{C}\cosh \left(2 \sqrt{C}x^-\right)\mp\sinh \left(2 \sqrt{C}x^-\right)\right)\right)\partial_{x^-}\nonumber\\
&+\left(\alpha_++\frac{1-f(\rho )^2}{8\sqrt{C}f(\rho )}\left(\alpha_-4 \sqrt{C} \cosh \left(2 \sqrt{C}x^-\right)\mp\sinh \left(2 \sqrt{C}x^-\right)\right)\right)\partial_{x^+}\, .
\end{align}
The differential operators \eqref{gendifBTZ} and \eqref{gendifBTZa} might not look like the standard basis for the local Killing vectors on BTZ. Nevertheless, they locally satisfy the Killing equation for \eqref{BTZmet1} and the expected $sl(2,\RR)_L\times sl(2,\RR)_R$ algebra.

Having evidence that the state  $ | \Sigma_{\text{cross}}\rangle$ is a natural probe (with usual geometric properties we associate to BTZ), we can infer from \eqref{UstateBTZ} that  
\begin{align}\label{UstateBTZe}
 &| U(x)\rangle_{\rm BTZ}=\cr
 &  e^{-\frac{i}{4}\left((8C-2)\ell_0-(1+4C)(\ell_{1}+\ell_{-1})\right)x^{+}}\,e^{-\frac{i}{4}\left((8C-2)\bell_0-(1+4C)(\bell_{1}+\bell_{-1})\right)x^{-}}e^{-\frac{\rho}{2} (\ell_1-\ell_{-1} +\bell_1-\bell_{-1} )}  | \Sigma_{\text{cross}}\rangle\,.
\end{align}
One can obtain $| U(x)\rangle_{\rm BTZ}$ from the gauge transformation that relates global AdS$_3$ and BTZ, and using \eqref{Ustateglobale}. We found, however, instructive to take a perspective where the metric is the first input and from there build \eqref{UstateBTZe}.

 The Ishibashi state $ | \Sigma_{\text{Ish}}\rangle$ also leads to barred differential operators. Acting on \eqref{innerbasisBTZb} with \eqref{eq:darot}, we get barred differential operators similar to those in \eqref{gendifBTZa}, but with an overall minus sign in $\mathcal{\bar L}_{\pm}$. These differential operators are still Killing vectors and they follow the $sl(2,\RR)_L\times sl(2,\RR)_R$ algebra by definition.


We now return to building $\Phi_{k, \bar k}(x)$. To start consider \eqref{hwcond}: given \eqref{gendifBTZ}, it is clear that $\Phi_{0, 0}(x)$ is non-separable in any of its variables, which makes \eqref{hwcond} very difficult to solve.  In order to simplify \eqref{hwcond}, we will make a change of variables; using \eqref{ToPoincare} 
we now have
  \begin{align}\label{gendifPoin}
\mathcal{L}_{0}  &=-\frac{i \left(r_+^2+1\right) Z^2 \partial_{X^+}+Z \left(r_+^2 (X^-+1)+X^--1\right) \partial_{Z}+\left(r_+^2 (X^-+1)^2+(X^--1)^2\right) \partial_{X^-}}{4 r_+}\, ,\nonumber\\
\mathcal{L}_{1} &=-\frac{i (r_+-i)^2 Z^2 \partial_{X^+}+( (r_+-i) X^-+r_++i) \left(( (r_+-i) X^-+r_++i) \partial_{X^-}+(r_+-i) Z \partial_{Z}\right)}{4 r_+}\,,\nonumber\\
\mathcal{L}_{-1} &=-\frac{i (r_++i)^2 Z^2 \partial_{X^+}+( (r_++i) X^-+r_+-i) \left(( (r_++i) X^-+r_+-i) \partial_{X^-}+(r_++i) Z \partial_{Z}\right)}{4 r_+}\, .
\end{align}
The barred operators are defined analogously with $X^+\leftrightarrow X^-$.  The advantage of  \eqref{gendifPoin}, relative to \eqref{gendifBTZ}, is that the differential operators just involve powers on the coordinates, and hence we can find a suitable polynomial solution to \eqref{hwcond}.  The unique solution to \eqref{hwcond} reads
 \begin{align}\label{hwBTZ}
 \Phi_{0,0}(x)&=Z^{-2 h} \left(-1+\frac{\left(X^-+\frac{r_++i}{r_+-i}\right) \left(X^++\frac{r_++i}{r_+-i}\right)}{Z^2}\right)^{-2 h}\\\nonumber
&=\left(2 r \left(r_+^2+1\right) \cosh (r_+ \phi )+\sqrt{r^2-r_+^2} \left((r_+-i)^2 e^{-r_+ t}+(r_++i)^2 e^{r_+ t}\right)\right)^{-2h}
 \end{align}
where in the second line we have changed to BTZ coordinates in \eqref{BTZmet}. And as expected the solution \eqref{hwBTZ} not separable in this coordinate system. Acting with $\mathcal{L}_{-1}(x)$, and $\mathcal{\bar L}_{-1}(x)$ in \eqref{hwBTZ}, and inspired by the the Jacobi polynomial form of the global case \eqref{AdSfield}, the general expression for a descendant of \eqref{hwBTZ} reads
\begin{align}\label{BTZfieldour}
 \Phi_{k,\bar k}(x)={C_{k,\bar k}}&\left(\frac{Z}{{(X^-+a) (X^++a)-Z^2}}\right)^{2 h} \left(a \frac{ (X^-+a) (X^++\frac{1}{a})- Z^2}{(X^-+a) (X^++a)-Z^2}\right)^{\bar k}\left(a\frac{(X^-+\frac{1}{a}) (X^++\frac{1}{a})-Z^2}{(X^-+a) (X^++\frac{1}{a})- Z^2}\right)^k \nonumber\\
&{\left(a^2-1\right)^{2h}} P_k^{(\bar k-k,\, 2h -1)}\left(1- 2 \frac{ (X^-+a) (X^++\frac{1}{a})- Z^2}{(X^-+a) (X^++a)-Z^2}\cdot\frac{(X^-+\frac{1}{a})) (X^++a)- Z^2}{(X^-+\frac{1}{a}) (X^++\frac{1}{a})-Z^2}\right)\,,
\end{align}
where $a\equiv \frac{i+r_+}{-i+r_+}$, and $C_{k,\bar k}$ is same factor as in \eqref{AdSfield}.  It is straight forward to verify that $ \Phi_{k,\bar k}(x)$ in \eqref{BTZfieldour} satisfies 
the d'Alembertian equation on the static BTZ background. 

Having an explicit expression for $ \Phi_{k,\bar k}(x)$, we can compute the overlap of two states \eqref{changeb} for the BTZ black hole. Using \eqref{BTZfieldour}, we see that he sum we need to perform in \eqref{innergen} is exactly equal to \eqref{mastersum2}, where
\begin{align}
&X= \frac{ |\tau_i|}{ |\gamma_i|}~,\qquad Y= \frac{ |\tau_f|}{ |\gamma_f|}~,\qquad r=|a|^2\sqrt{\frac{\tau_f \gamma_f^*}{\tau_f^* \gamma_f}\frac{\tau_i^*\gamma_i}{\tau_i \gamma_i^*}}~,\qquad s=|a|^2\sqrt{\frac{\tau_f ^*\gamma_f^*}{\tau_f \gamma_f}\frac{\tau_i\gamma_i}{\tau_i ^*\gamma_i^*}}~,
\end{align}
and
\be
\gamma_{i,f}\equiv(X^-_{f,i}+a)(X_{f,i}^++a)- Z_{f,i}^2~,\qquad \tau_{i,f}\equiv(X^-_{f,i}+\frac{1}{a}) (X_{f,i}^++a)- Z_{f,i}^2~.
\ee
Using the result for the sum \eqref{mastersum2}, with the previous definition for $X,Y,r$, and $s$, we find the the overlap of the two states in the BTZ black hole:
\be\label{resultBTZ}
\langle  U(x_f)  | U(x_i)\rangle= \frac{\left(\sigma(x_i,x_f)+\sqrt{\sigma^2(x_i,x_f)-1}\right)^{-(2h-1)}}{2\sqrt{\sigma^2(x_i,x_f)-1}}~,
\ee
where $\sigma(x_i,x_f)$ is the geodesic distance for Poincare \eqref{geodlengthpoin}, which can be rewritten as the geodesic length in BTZ \eqref{geodlengthr} using \eqref{fromPoin}. With no surprises, this is in complete agreement with \eqref{SCHlength}.

It is interesting to analyse the behaviour  of the field \eqref{BTZfieldour} in the BTZ coordinates. Looking at \eqref{fromPoin}, we see that the BTZ boundary $r\rightarrow\infty$ is located at $Z\rightarrow 0$, and in this limit we have  $\Phi_{k,\bar k}\rightarrow 0$. The horizon  $r=r_+$ is at the Poincare boundary $(X^+,X^-,Z)\rightarrow \infty$, where $\Phi_{k,\bar k}$ as well vanishes. This behaviour, together with the fact that solves the BTZ wave equation, shows that \eqref{BTZfieldour} behaves as a \emph{quasi-normal mode} for the black hole. However, it is not a traditional BTZ quasi-normal mode as those built in, e.g., \cite{Chan1997,Birmingham:2001hc,Cardoso:2001hn,Birmingham2002}. There are a few discrepancies, and a few similarities, with this literature that are worth highlighting.
\begin{enumerate}
\item {\it Highest weight condition.} As it was observed in \cite{Chen2010,Zhang2011}, imposing the highest weight conditions \eqref{raiseAdSfield}-\eqref{hwcond} leads to eigenfunctions that obey the quasinormal modes conditions. This is a first indication that $\Phi_{k, \bar k}(x)$ should have been regular throughout, as they certainly are.  
\item {\it Separability of eigenfunctions.} The most canonical way to find solutions to the Casimir equation \eqref{casimir} is by casting the basis of solutions in a Fourier decomposition in $(t,\phi)$, which are the natural directions for the Killing symmetries of the black hole. This leads a eigenfunctions that are separable functions in the coordinate system $(r,t,\phi)$, in strike contrast to  \eqref{BTZfieldour}. The construction of the operators ${\cal L}_a$ in \cite{Chen2010}, which is used to build a basis for quasinormal modes, is as well compatible with the separability ansatz. From a technical point of view, our lack of separability could be attributed to the unitary condition we enforce in \eqref{UstateBTZe}:  this leads to a group elements that are simply different to those used in  prior work.\footnote{We could have parametrized the group elements in \eqref{UstateBTZe} so that we obtain the same basis for ${\cal L}_a$ in \cite{Chen2010} that leads to separability. However, with this choice the state is not unitary and hence $\langle U| \neq (|U\rangle)^\dagger$.}
\item {\it Periodicity conditions.} By design, the connections $(A,\bar A)$ that characterize BTZ in the Chern-Simons formulation  have the following feature \cite{MaxTesis,Gutperle:2011kf}: they are single valued along the thermal cycle in Euclidean signature (smoothness of the Euclidean cigar geometry), and carry a non-trivial holonomy around the spatial cycle (an indication that the connection has a finite size horizon). This is reflected in \eqref{BTZfieldour} by the fact that our eigenfunctions are not periodic as we take $\phi\sim \phi+2\pi$, but are periodic under $t\sim t+i2\pi/r_+$. This is clearly not a feature of the modes built in \cite{Chan1997,Birmingham:2001hc,Cardoso:2001hn,Birmingham2002}, which are decomposed in periodic Fourier modes along the $\phi$ direction. 
\item {\it Inner Product.} Despite the two differences above, it is interesting to note that if we evaluated the overlap \eqref{innergen} using the quasinormal modes in \cite{Birmingham:2001hc}, it would lead to \eqref{resultBTZ}.  The derivations are shown in appendix \ref{sec:quasi}. This indicates that the bulk-to-bulk correlation functions are not sensitive to how we represent $\Phi_{k,\bar k}(x)$. 
\end{enumerate}

\section{CFT interpretation}\label{sec:cft}

Here we discuss the CFT interpretation of the results above. In particular, consider computing a Wilson line in AdS$_3$, ending at the AdS boundary at the two boundary points $z_1, z_2$ at radial coordinate $\rho_1, \rho_2$ with generic boundary conditions $U_1$, $U_2$ at each endpoint. What, precisely, is this object in the CFT? 

The considerations of the previous section should make it clear that the resulting object is a suitably smeared two-point function, and here we simply provide a purely boundary interpretation of this smearing procedure. The kinematics of these procedure are very familiar from the language of the the HKLL construction \cite{Hamilton:2006az,Hamilton:2005ju} and this section may be understood as a translation of some of those results into the language of Chern-Simons gravity. Since $|\Sigma_{\rm cross}\rangle$ leads to the standard conventions in the metric formulation, relative to $|\Sigma_{\rm Ish}\rangle$,  we will focus on the role of the crosscap Ishibashi state in this section.

Let us first consider what data we have; at each endpoint $z_i$ in the CFT we are supplied with a length scale $e^{-\rho_i}$ arising from the cutoff and an $\slt$ element arising from the boundary conditions on the Wilson line. There is a natural way to associate this data with the global descendants of a boundary operator $\sO$: first, act on the Ishibashi crosscap state $|\Sigma_{\rm cross}\rangle$ with the $\slt$ elements $U_{i}$ as explained in detail in Section \ref{sec:hilbert} to construct a state $|U_i\rangle$, where the states in the highest weight representation are understood as conformal descendants of $\sO$. Next, remove two discs from the CFT, each centered at $z_i$ with radius $e^{-\rho_i}$; on each of these discs place the boundary data appropriate to $|U_i\rangle$. This is the CFT dual to the open-ended Wilson line with boundary condition $U_i$. 

Mathematically this is essentially the same construction as \cite{Verlinde:2015qfa,Miyaji:2015fia,NakayamaOoguri2015}. There are two main differences: in all of these works the specification of the $\slt$ element was interpreted to specify a point in AdS$_3$ rather than a boundary condition on a Wilson line. Furthermore in \cite{Verlinde:2015qfa,NakayamaOoguri2016,LewkowyczTuriaciVerlinde2017} the full Virasoro group was considered rather than just its global subgroup. The former is just a matter of interpretation, and we will touch briefly on the latter in the conclusion. 

\subsection{Example: CFT on the plane} 
We now present some elementary computations to explain how this works in the basic case of Poincar\'e AdS$_3$ in coordinates:
\be
ds^2 = d\rho^2 + e^{2\rho} dz d\bz~, \label{poinc} 
\ee
dual to the CFT on the plane with complex coordinates $z, \bz$. Rather than working with boundary data on the edge of an excised disc at each endpoint, it is more convenient to perform the state-operator correspondence to map each descendant on the edge to a local operator at the center of the disc. As there are an infinite number of states in the sum, this is a very non-local operator which we denote by $\sO_{U_i}(z_i, \bz_i)$. We will use a variant of the HKLL construction to compute the two-point function
\be
\langle \sO_{U_1}(z_1, \bz_1) \sO_{U_2}(z_2,\bz_2) \rangle~, \label{Ustate2pt} 
\ee
and then reproduce this answer from a Wilson line computation. 

Focus on the first endpoint at $(z_1, \bz_1)$. We first consider the case where the boundary state $U_1$ is the crosscap Ishibashi state $|\Sig_{\rm cross}\rangle$ itself. Consider the disc centered at $z_1$ in the CFT, with radius $e^{-\rho}$; we would like to place the boundary data corresponding to the crosscap Ishibashi state:
\be
|\Sig_{\rm cross}\rangle = \sum_{m} (-1)^{m} |m,m\rangle = \sum_{m} (-1)^{m} c_m^2 \ell^m_{-1} \bar{\ell}^m_{-1} |h, h\rangle~,
\ee
where the normalization constant in each sector is $c_m = \sqrt{\frac{\Gamma(2h)}{\Gamma(m+1)\Gamma(m+2h)}}$. We now use the state-operator correspondence to replace each state on the disc $\ell^m_{-1} \bar{\ell}^m_{-1} |0\rangle$ with the operator $\p^m \bar{\p}^m \sO(z_1)$ at the center. However, we should note that the evolution from the center of the disc to the edge will cause each state's amplitude to be multiplied by a factor of $e^{+\rho(2h + 2m)}$.  Compensating for this, the operator that creates the crosscap state on a disc of radius $e^{-\rho_1}$ is
\be
|\Sig_{\rm cross} \rangle \to  \sO_{\Sig}^{(\rho_1)}(z_1,\bz_1) = \sum_{m} c_m^2 (-1)^{m} e^{-\rho_1(2h + 2m)} \p^m \bar{\p}^m \sO(z_1,\bz_1)~.
\ee
The sum over derivatives of the local operator $\sO$ can be written as an integral over the following kernel \cite{Goto:2017olq,Anand:2017dav} 
\begin{align}
\sO_\Sig^{(\rho_1)}(z_1,\bz_1) & = \frac{2h-1}{\pi} \int dz d \bz \le(\frac{e^{-2\rho_1} - (z - z_1) (\bz - \bz_1)}{e^{-\rho_1}}\ri)^{2h-2} \sO(i z, i\bz) \cr
& \equiv K(\rho_1, z_1,\bz_1)[\sO]~, \label{Knot} 
\end{align}
where in the last line we have introduced some new notation. We note that this is nothing but the usual HKLL smearing kernel in Euclidean signature, though our interpretation is somewhat different. 

We now consider deforming away from the crosscap state to a more general $U$-state. The $\slt$ generators have a simple geometric action on the plane, and this geometric action results in a transformation of the parameters in  kernel $K$. In particular, if we parametrize the $\slt$ element $U_1$ in a convenient way as
\be
U_1 = e^{-w_1\frac{i}{2}(-2 L_0 - L_1 - L_{-1})}e^{\mu_1(L_{-1}-L_{1})}e^{\bw_1\frac{i}{2}(-2L_0 + (L_1 + L_{-1}))}~, \label{Uparam} 
\ee
then it is shown in Appendix \ref{app:CFTkernel} that the appropriate smeared operator is
\be
\sO^{(\rho_1)}_{U_1}(z_1, \bz_1) = K(\rho_1+\mu_1, z_1 + e^{-\rho_1} w, \bz_1 + e^{-\rho_1} \bw_1)[\sO] ~. \label{gensmear}
\ee
We now pause to interpret this result from the point of view of HKLL. Recall that the smearing function \eqref{gensmear} corresponds to the HKLL representation of a bulk field in Poincar\'e coordinates \eqref{poinc},
where the precise coordinate values of the bulk reconstructed field are
\be
(\rho_a, z_a, \bz_a) = (\rho_1 + \mu_1, z_1 + e^{\rho_1} w_1, \bz_1 + e^{-\rho_1} \bw_1) ~,\label{poinval} 
\ee
In particular, the proper distance $D$ within AdS$_3$ between two points with coordinate values $(\rho_a, z_a, \bz_a)$ and $(\rho_b, z_b, \bz_b)$ satisfies $\cosh D = \sig$ where
\be
\sig = \frac{e^{-2\rho_a} + e^{-2 \rho_b} + (z_a - z_b)(\bz_a - \bz_b)}{2 e^{-\rho_a - \rho_b}}~. \label{2pt}
\ee
Let us now consider computing the two-point function \eqref{Ustate2pt} of two $U$ states inserted at distinct points on the boundary $z_1, z_2$. This is a well-posed CFT computation involving integrals over two $K$ kernels. Rather than repeat it here, we simply note that it is a standard HKLL computation, and by construction the result is the usual bulk-to-bulk AdS$_3$ propagator between points with Poincare coordinate values given by \eqref{poinval} (and a corresponding relation relating $(\rho_a, z_a, \bz_a)$ to $(\rho_2, z_2, \bz_2)$), i.e.
\be
\langle \sO^{(\rho_1)}_{U_1}(z_1, \bz_1) \sO^{(\rho_2)}_{U_2}(z_1,\bz_1)\rangle = \frac{e^{-2h D}}{1-e^{-2D}}~.
\ee
Such expressions are by now very familiar. 

We will now reproduce this CFT result from a Chern-Simons computation. In particular, we consider the following gauge connections for Poincar\'e AdS$_3$ in Euclidean signature:
\be
a = -\frac{i}{2}\le(-2 L_0 - L_{1} - L_{-1})\ri) dz~, \qquad \bar{a} = -\frac{i}{2}\le(-2 L_0 + L_{1} + L_{-1}\ri) d\bz~, \qquad b = e^{(L_{-1} - L_{1})\frac{\rho}{2}}~. \label{ads3conneuc} 
\ee
As usual the full connections are related to the objects recorded here by $A = b^{-1}\le(a + d\ri)b$, $\bA = b(\bar{a} + d)b^{-1}$. Full conventions are in Appendix \ref{app:cs}; in particular these connections are equivalent to those in \eqref{sl2conn2} with $C \to 0$, together with the usual Euclidean continuation $x^+ \to z, x^{-} \to -\bz$ and a rescaling of the field-theory directions by a factor of $2$; the last step is convenient so that the resulting coordinate system is precisely equivalent to \eqref{poinc}. 

The prescription above states that the two point-function \eqref{Ustate2pt} is calculated in the Chern-Simosn representation by the following matrix element:
\be
\langle \sO^{(\rho_1)}_{U_1}(z_1, \bz_1) \sO^{(\rho_2)}_{U_2}(z_1,\bz_1)\rangle = \langle U_2 |G\left(\mathcal{P}\,e^{-\int_\gamma {A}}\right) \bar G\left(\,\mathcal{P}\,e^{-\int_\gamma {\bar A}}\,\right)| U_1\rangle ~. \label{toconfirm} 
\ee
We may easily verify this relation. \eqref{genans} now tells us that the right-hand side of this expression is equal to this matrix element is equal to 
\be
\langle U_2 |G\left(\mathcal{P}\,e^{-\int_\gamma {A}}\right) \bar G\left(\,\mathcal{P}\,e^{-\int_\gamma {\bar A}}\,\right)| U_1\rangle  = \frac{e^{-h \al}}{1-e^{-\al}}~, 
\ee
where as usual $\al$ is defined as the $L_0$ conjugacy class of the following group element: 
\be
V e^{-\al L_0} V^{-1} = \tilde{g}_R(x_2) U_2^{-1} g_L(x_2) g_L(x_1)^{-1} U_1 \tilde{g}_R(x_1)^{-1} ~,
\ee
with
\be
\tilde{g}_{R}(x) = e^{\rho\frac{L_1 - L_{-1}}{2}} e^{-a_{z} z} \qquad g_L(x) = e^{\bar{a}_z \bz}e^{\rho\frac{L_1 - L_{-1}}{2}}~.
\ee
Computing $\al$ from here and the explicit representation of $U_{1,2}$ as in \eqref{Uparam}, we find that it is  equal to $2 D$ as defined above \eqref{2pt}; thus we find that the Chern-Simons computation agrees with the CFT result, confirming \eqref{toconfirm}. 

 Note that everything in this computation is fixed by kinematics, and we have simply shown how the $\slt$ parameters characterizing the boundary conditions combines with the geometric data to give the familiar HKLL result. 

%
%

 
 

\section{Discussion}\label{sec:discussion}

We provided a full quantum mechanical description treatment of worldline degree of freedom of a Wilson line in $SO(2,2)$ Chern-Simons theory. This degree of freedom allowed us to build a local probe in the Chern-Simons description of AdS$_3$ gravity. There are a few striking features of this probe which we highlight. 

\begin{enumerate}
\item We designed states in  the worldline quantum mechanics such that they would couple to both $(A,\bar A)$. This condition naturally introduced the notion of \emph{rotated Ishibashi states}, which we denote as $|U\rangle$, and their coupling to the connections creates a background spacetime metric 
\be\label{eq:g3}
g_{\mu\nu}={1\over 2} {\rm Tr}(A_\mu -\tilde \bA_\mu )(A_\nu -\tilde \bA_\nu)~,
\ee
where $\tilde\bA=U^{-1}\bar A U $.  These rotated Ishibashi states are at the core of giving   a geometric, and hence local, interpretation to  $W_{\mathcal{R}}(x_f, x_i)$. In particular, we showed that $W_{\mathcal{R}}(x_f, x_i)$ is the bulk-to-bulk propagator of a scalar field propagating on \eqref{eq:g3}. The most natural choice of rotated state that leads to regular background metrics is the crosscap state \eqref{cross}, i.e. $|U\rangle=|\Sigma_{\rm cross}\rangle$.
\item Using purely the Chern-Simons formulation, we can build local bulk fields that probe the background geometry \eqref{eq:g3}. These local probes are defined in \eqref{innerbasis} and we investigated some their properties for global AdS$_3$ and the static BTZ black hole. 
\end{enumerate}

It is very satisfactory that our choice $|U\rangle=|\Sigma_{\rm cross}\rangle$ is compatible with the proposals in \cite{NakayamaOoguri2015,Miyaji:2015fia,Verlinde:2015qfa}, and we also reproduce the smearing functions of the HKKL \cite{Hamilton:2005ju,Hamilton:2006az} proposal for vacuum solutions. 
  This is expected since the symmetries of AdS$_3$ constrain heavily the resulting bulk field, leaving little room for disagreement at this level of discussion. Perhaps the interesting difference of our approach is that our construction leaves room to consider other probes $|U\rangle$, and highlights some of the gauge dependence in the construction of the local field $\Phi_{k,\bar k}(x)$, which we emphasised around \eqref{Ustate22}.  For black holes the situation is more delicate: for instance, it would be interesting to compare and complement the proposals in \cite{HamiltonKabatLifschytzEtAl2007,PapadodimasRaju2014,GuicaJafferis2017,Goto:2017olq, daCunha:2016crm} with our derivations in Sec. \ref{sec:BTZ}.  Along these lines it would be interesting to carry out our derivations for the rotating BTZ black holes, and other backgrounds in 3D gravity we have not explored. 
 
We comment very briefly on one other aspect; as we have been able to reproduce bulk-to-bulk propagators from the Chern-Simons description of 3d gravity, it is worth wondering whether all of the aspects of the quantum field theory of a scalar field on a gravitational background can be obtained from the Chern-Simons computation, e.g. can we obtain a one-loop scalar field determinant on a BTZ black hole background? As this is essentially the same information as the bulk-to-bulk propagator, we might think so. Indeed we expect the logarithm of the one-loop determinant $\mathcal{W}$ to be the sum over connected Feynman diagrams, which in our context is the sum of Wilson lines that each wrap the horizon $n$ times on topologically distinct paths $C_{n}$. We find:
\be
\mathcal{W} = 2\sum_{n = 1}^{\infty} \frac{1}{n} \Tr_{\mathcal{\sR}} \le[{\cal P} \exp\le(- \oint_{C_n} A\ri){\cal P} \exp\le(-\oint_{C_n} \bar A\ri)\ri] = 2\sum_n \frac{1}{n} \le(\frac{e^{-h n \al}}{1- e^{-n\al}}\ri)^2, 
\ee
Here we have assumed that the topologically trivial path does not contribute; the factor of $2$ arises from positive and negative $n$. The combinatoric factor $\frac{1}{n}$ is a symmetry factor\footnote{To understand this symmetry factor, consider having $n$ objects which must be connected into closed paths; there are $(n-1)!$ distinct cycles (and not $n!$) as the starting point of the cycle is arbitrary. Compensating for this overcounting results in this factor of $n$.}  and as usual $\al$ is the conjugacy class of the holonomy of $A$ (or $\bar{A}$) around the black hole; on the BTZ background it evaluates to $\al = 2\pi r_+$. The result above is then precisely the logarithm of the usual one-loop scalar determinant on a black-hole background; see e.g. \cite{Giombi:2008vd} for details and a repackaging of this result in CFT language. 

An important issue that we have not addressed is quantum corrections due to fluctuations of the background connections. This would capture $1/c$ corrections, i.e. corrections controlled by the AdS radius in Planck units, or equivalently subleading terms controlled by the level of the Chern-Simons theory.  Work in this direction has been done for $SL(2)$ Chern-Simons theory, where Virasoro conformal blocks are known to be tied to appropriate Wilson line in Chern-Simons \cite{Verlinde1990,Elitzur1989,Witten:1988hf}. Recent developments for this holomorphic theory include \cite{Alkalaev2015,Hijano2015a,Bhatta:2016hpz,Alkalaev2016,Besken2016,Fitzpatrick2017}. It would be interesting to evaluate  $1/c$ corrections of our worldline quantum mechanics;  in this case  we expect that the intertwining of the two copies of $sl(2)$ will produce interesting features. For example, we should be able to probe if the global conditions in \eqref{Ishieq1} are enhanced to the Virasoro conditions on the Ishibashi state \cite{NakayamaOoguri2016,LewkowyczTuriaciVerlinde2017}, or something completely different, such as the conditions proposed in \cite{Anand:2017dav}. We leave these questions for future work. 

Another natural direction forward is to use our construction to build probes in $SL(N)\times SL(N)$ Chern-Simons theory. This would provide a unique way to build local probes in higher spin gravity. A discussion of Ishibashi states for ${\cal W}_3$ algebra was done in \cite{NakayamaSuzuki2018}, which is a natural starting point for future investigations. 





\section*{Acknowledgements}

It is a pleasure to thank M. Guica, D. Hofman, H. Ooguri, P. Sabella-Garnier, H. Verlinde, and C. Zukowski. AC's and NI's work was performed in part at the Aspen Center for Physics, which is supported by National Science Foundation grant PHY-1607611. AC would like to thank Tsinghua Sanya International Mathematics Forum, Mainz Institute for Theoretical Physics (MITP),  and Universidad Catolica of Chile for their hospitality and partial support during the completion of this work. NI would like to thank Delta ITP at the University of Amsterdam for hospitality during the completion of this work. AC and EL are supported by Nederlandse Organisatie voor Wetenschappelijk Onderzoek (NWO) via a Vidi grant. NI is supported in part by the STFC under consolidated grant ST/L000407/1. This work is part of the Delta ITP consortium, a program of the NWO that is funded by the Dutch Ministry of Education, Culture and Science (OCW).

\appendix

\section{Properties of $so(2,2)$ representations}\label{app:sl2}

In this appendix we collect a set of definitions, conventions and identities that are relevant for our manipulations in the highest weight representation, and the rotated Ishibashi states. 

\subsection{$sl(2,\mathbb{R})$ conventions}
The Lie algebra for $sl(2,\mathbb{R})$ is given by
 \begin{align}\label{sl2alg}
     [L_0,L_{\pm}] = \mp L_{\pm}\,,\qquad [L_1,L_{-1}] = 2L_{0}\,,
\end{align}
Our conventions for the fundamental representation of  $sl(2,\mathbb{R})$ is
 \begin{equation}\label{sl2fund}
L_0 = \begin{pmatrix}
    1/2  & 0    \\
      0 &  -1/2
\end{pmatrix}\,, \quad
L_1 =  \begin{pmatrix}
    0  & 0    \\
      -1 &  0
\end{pmatrix}\,, \quad
L_{-1} =  \begin{pmatrix}
    0  & 1    \\
      0 & 0 
\end{pmatrix}\,.
\end{equation}
In our conventions, the Lie algebra metric reads
\be\label{sl2metric}
\eta_{00}=\frac{1}{2}\,,\qquad \eta_{+-}=\eta_{-+}=-1~.
\ee

 For the highest weight representation of $sl(2,\mathbb{R})\times sl(2,\mathbb{R})$, we denote the generators as $(\ell_a, \bar\ell_a)$. Some of the basic relations we use in the main text are 
 \begin{align} \label{norm}
\ell_{0} |h,k,\bar k\rangle &= (h+k)\,|h,k,\bar k\rangle~ , \cr
\ell_{-1} |h,k,\bar k\rangle &= \sqrt{(k+1)(k+2h)} |h, k+1,\bar k\rangle~, \\
\ell_1 |h,k,\bar k\rangle &= \sqrt{k(k+2h-1)}|h,k-1,\bar k\rangle~,  \nonumber
\end{align}
 where $k=0,...,\infty$. The barred operators $\bell_a$ act analogously as \eqref{norm}, but in the states labeled by $\bar k$.

\subsection{Completeness of rotated Ishibashi states} \label{app:completeness}

Here we establish a completeness relation for the $|U\rangle$ states:
\be
\int dU  |U \rangle \langle U | = \frac{(2\pi)^2}{2(2h-1)} \mathds{1} ~.\label{completenessapp} 
\ee
We note that if the right-hand side exists, it must be equal to a multiple of the identity by $\slt$ invariance; thus the only question is whether or not the integral converges, and what the normalization factor is if it does. As the group is non-compact the convergence is not (to our knowledge) actually guaranteed. Thus we perform an explicit computation in coordinates. In particular we view the $\slt$ group manifold as global AdS$_3$ and place on it the usual global coordinates $(\rho, t, \phi)$. It is important to note that we work here not $\slt$ and not with its universal cover, and thus {\it both} coordinates $t$ and $\phi$ are periodic with period $2\pi$. 

The explicit matrix elements between the $|U(\rho, t, \phi)\rangle$ states and the discrete highest-weight states $|h,k,\bar{k}\rangle$ can be constructed via the usual methods of finding the highest-weight state and systematically acting with the raising operators. The result is precisely that given in a (slightly) different context in the bulk of the paper \eqref{AdSfield}:
\be
\langle U(\rho,t,\phi) | h; k, \bar{k}\rangle = C_{k,\bar k}\, e^{-2iht}e^{-it(k + \bar{k})-i\phi(k - \bar{k})}(\tanh \rho)^{\bar k- k}(\cosh \rho)^{-2h}P_k^{(\bar k-k,\, 2h -1)}(1-2\tanh^2\rho)\,,
\ee
where $P_n^{(a,\, b)}$ are Jacobi polynomials, and $C_{k,\bar k}\equiv(-1)^k\sqrt{\frac{k! (2 h+\bar k-1)!}{\bar k! (2 h+k-1)!}}$. Note also that in these coordinates on the group manifold the Haar measure is just the usual volume element on AdS$_3$, i.e.
\be
\int dU = \int d\rho dt d\phi \sinh\rho \cosh\rho~.
\ee
With this in hand, we simply directly compute the following matrix elements:
\be
I_{m,\bar{m};k,\bar{k}} \equiv \int dU \langle h; m, \bar{m}|U \rangle \langle U | h; k, \bar{k}\rangle~.
\ee
From the matrix elements above, we see that this integral is proportional to $e^{-it(k+\bar{k} - m-\bar{m})} e^{-i\phi(k - \bar{k} - m + \bar{m})}$; thus the integrals over $t$ and $\phi$ result in a vanishing matrix element unless $k = m$ and $\bar{m} = \bar{k}$. We conclude then that
\be
I_{m,\bar{m};k,\bar{k}} = \delta_{m,k} \delta_{\bar{m},\bar{k}} N_{k,\bar{k}}~.
\ee
The normalization factor is given by
\be
N_{k,\bar{k}} = (2\pi)^2 \int d\rho \sinh\rho \cosh\rho \le((\tanh \rho)^{\bar k- k}(\cosh \rho)^{-2h}P_k^{(\bar k-k,\, 2h -1)}(1-2\tanh^2\rho)\ri)^2~,
\ee
This is difficult to evaluate for generic $k$. However by $\slt$ invariance it must be independent of $k,\bar{k}$ (a fact we have also checked directly by numerical evaluation of the integral), allowing the integral to be performed for $k = \bar{k} = 0$, resulting in 
\be
N_{k,\bar{k}} = N_{0,0} = \frac{(2\pi)^2}{2(2h-1)}~.
\ee 
Assembling the pieces we find
\be
I_{m,\bar{m};k,\bar{k}} = \frac{(2\pi)^2}{2(2h-1)}\delta_{m,k}\delta_{\bar{m},\bar{k}}~,
\ee
which is precisely the completeness relation \eqref{completenessapp} that we set out to show. 

\section{Chern-Simons formulation of AdS$_3$ gravity}\label{app:cs}

With the purpose of setting up conventions, in this appendix we give a very short review of the Chern-Simons formulation of AdS$_3$ gravity. We refer the reader to the original articles \cite{Achucarro:1987vz,Witten:1988hc} and more recently in, e.g., \cite{Banados:1998gg,Ammon:2012wc} for further details.

The relevant Chern-Simons gauge group for AdS$_3$ gravity is $G=SO(2,2)$. The Einstein-Hilbert action can be written as
\bea
S_{\rm EH} [e,\omega]&=& S_{CS}[{\cal A}] \cr &=& \frac{k}{4\pi}\int_{\cal M} \Tr\left({\cal A} \wedge d{\cal A} + \frac{2}{3} {\cal A} \wedge {\cal A} \wedge {\cal A}\right)~,
\eea
with ${\cal A}\in  so(2,2)$.  
Here $k$ is the level of the Chern-Simons theory. The relation to the conventional gravitational vielbein and spin connection is
\be
{\cal A}_i= e^a_i P_a + \omega^a_i M_a~,
\ee
 where $M_a$ are Lorentz generators and $P_a$ are  translations in $so(2,2)$. 
  
 It will be convenient to write the gauge group $SO(2,2)$ as $SL(2,\RR)\times SL(2,\RR)$. The flat connection ${\cal A}$ can then be decomposed as two pairs of connections 
\be
A= (\omega^a +{1\over \ell} e^a)L_a ~,\quad  \bar A= (\omega^a -{1\over \ell}e^a)\bar L_a~, \label{vbspin}
\ee
with $L_a={1\over 2}(M_a+\ell P_a)$, and  $\bar L_a={1\over 2}(M_a-\ell P_a)$. Here $\ell$ is the AdS radius, which for most of our work we will set $\ell=1$, and Newton's constant is related to the Chern-Simons level via
  \be\label{eq:level}
  k={\ell\over 4G_3}~.
  \ee
We will  denote the generators of $sl(2,\mathbb{R})$ simply as $L_a$. After performing this decomposition the action can be written 
\be
S_{\rm EH} = S_{CS}[A] - S_{CS}[\bA]~,
\ee
where the trace operation used in defining the Chern-Simons form is now the usual bilinear form on the $sl(2,\mathbb{R})$ Lie algebra. 

\subsection{Metrics, connections, and geodesic distances}\label{sec:coord}

In this appendix we gather various properties used for global AdS and the BTZ background. We present the relevant information in Chern-Simons formulation, and the metric formulation.  For the later, we gather the different coordinate systems used and the relevant geodesic distances. \\
\\
In Chern-Simons formulation, we write the pair of $sl(2,\mathbb{R})\times sl(2,\mathbb{R})$ as
\be
 A(x) = g_L(x) d g_L(x)^{-1}\,, \quad  \tilde \bA(x) = \tilde g_R(x)^{-1} d  \tilde g_R(x)\,, 
\ee
In this section we add the tilde in the right sector, for consistency with the conventions used in the main text. When the connections are constant in boundary coordinates, we can cast the group elements as  
 \be\label{glgr}
 g_L(x)= b(\rho)^{-1}e^{-a_{\mu}y^{\mu}}\,,\quad  \tilde g_R(x)=e^{\bar a_{\mu}y^{\mu}}b(\rho) ^{-1}\, \qquad y^\mu =(t,\phi)~,
 \ee
 where $b(\rho)$ parametrizes the choice of radial variable, and $a_\mu$, and $\bar a_\mu$ are constant elements of the $sl(2)$ algebra. We can use the previous reparametrization to express the BTZ and global AdS metric, with\footnote{We chose these explicit form of the connections because they result into unitary group elements \eqref{glgr} when we consider the highest weight representation with $(\ell_n)^{\dagger}=\ell_{-n}$. This is required by the purposes of the main text. For readers familiar with the previous literature in 3d gravity in CS formalism, it will be comforting to know that \eqref{sl2conn2} is related to the more familiar form of the BTZ connections:
  \begin{align}\nonumber
a= \left(L^e_+-C\,L^e_-\right)dx^+~, \quad\bar{a}=-\left(L^e_--C\,L^e_+\right)dx^-~,\qquad b(\rho)=\exp(\rho L^e_0)\,.
\end{align}
  via the following automorphisms: 
 \begin{align}\nonumber
L:\qquad L^e_1= i(2 L_0+L_1+ L_{-1})/4\,,\qquad L^e_{-1}= 2 i L_0-i L_1-i L_{-1}\,,\qquad L^e_0=-(L_1-L_{-1})/2\,,\\
R:\qquad L^e_1=2 i L_0+i L_1+i L_{-1}\,,\qquad L^e_{-1}= i(2 L_0-L_1- L_{-1})/4\,,\qquad L^e_0=-(L_1-L_{-1})/2\,,\nonumber
\end{align}
The automorphism labelled by $R$ is performed in the right sector, and analogously for the left sector. }
%
\begin{align}\label{sl2conn2}
&a=-\frac{i}{4}\left((8C-2) L_{0} -(1+4C)(L_{1}+L_{-1})\right)dx^+\,, \qquad  b({\rho})=e^{-(L_{1}-L_{-1})\rho/2}\,, \nonumber\\
&\bar a=\,\frac{i}{4}\left((8C-2)L_{0}+(1+4C) (L_{1}+L_{-1})\right)dx^-\,.
\end{align}
Here $\rho$ is the radial direction and $x^\pm =t\pm \phi$ with $\phi\sim \phi+2\pi$. Via \eqref{eq:effmetric}, these connections \eqref{sl2conn2} correspond to the metric:
\be\label{Banadosmet}
ds^2=d\rho^2-\frac{1}{4}(e^{\rho}-4 Ce^{-\rho})^2dt^2+\frac{1}{4}(e^{\rho}+4 Ce^{-\rho})^2d\phi^2\,.
\ee
%
For $C>0$, it is useful to define
\be
r=r_+\cosh^2(\rho-\rho_*)~,\qquad 4C= e^{2\rho_*}= r_+^2~,
\ee
which brings \eqref{Banadosmet} to the more familiar version of the (non-rotating) BTZ black hole:
\be\label{BTZmet}
ds^2=-(r^2-r_+^2)dt^2+\frac{dr^2}{r^2-r_+^2}+r^2d\phi^2\,,
\ee
For $C<0$ the background \eqref{Banadosmet} corresponds generically to a conical deficit. Setting $C=-1/4$, we recover from \eqref{Banadosmet} the global AdS$_3$ spacetime:
\be\label{AdSgl}
ds^2=-\cosh^2\rho \,dt^2+{d\rho^2}+\sinh^2\rho \,d\phi^2\,.
\ee
In the next sections, we will need as well that the geodesic distance between two spacelike separated points in the bulk $(x_f,x_i)$. For the metric in \eqref{Banadosmet}, the geodesic distance is  $D(x_f,x_i)=\operatorname{arcosh}\sigma(x_f,x_i)$, with
\begin{align}\label{geodlength}
\sigma(x_f,x_i)=& \cosh (\rho_f-\rho_i) \cosh \left(\sqrt{C} (\Delta t+\Delta \phi)\right) \cosh \left(\sqrt{C} (\Delta t-\Delta \phi)\right) \nonumber \\
&-\frac{1}{2}\left(4 C\, e^{-(\rho_f+\rho_i)}+\frac{e^{\rho_f+\rho_i}}{4 C}\right)\sinh \left(\sqrt{C} (\Delta t+\Delta \phi)\right) \sinh \left(\sqrt{C} (\Delta t-\Delta \phi)\right)\,.
\end{align}
which in the coordinates in \eqref{BTZmet} is :
\be\label{geodlengthr}
\sigma(x_f,x_i)=\frac{1}{r_+^2}\left( r_fr_i\cosh (r_+\Delta\phi )- \sqrt{(r_f^2-r_+^2)(r_i^2-r_+^2)}\cosh (r_+\Delta t)\right)\,.
\ee
Moreover, the BTZ metric is locally isomorphic to $AdS_3$ Poincare. Using the following coordinate change
\begin{align}\label{ToPoincare}
\frac{R^2-T^2+X^2+Z^2}{2 Z}=\frac{\sqrt{r^2-r_+^2} \sinh (r_+ t)}{r_+}\,,\quad\frac{R T}{Z}=\frac{r \cosh (r_+ \phi )}{r_+}\,,\\\nonumber
\frac{R^2+T^2-X^2-Z^2}{2 Z}=\frac{\sqrt{r^2-r_+^2} \cosh (r_+ t)}{r_+}\,,\quad\frac{R X}{Z}=\frac{r \sinh (r_+ \phi )}{r_+}\, ,
 \end{align}
the metric \eqref{BTZmet} becomes
\be\label{Poinmet}
ds^2=\frac{1}{Z^2}(-dT^2+dX^2+dZ^2)\,,
\ee
Null coordinates in this system are defined as $X^+=X+T$, and $X^+=X-T$. The geodesic distance is
\be\label{geodlengthpoin}
\sigma(x_f,x_i)=\frac{(T_f-T_i)^2+(X_f-X_i)^2+Z_f^2+Z_i^2}{2 Z_iZ_f}\,.
\ee
A solution of \eqref{ToPoincare}, in the quadrant where  $X,T,Z>0\,,1\geq T^2-X^2-Z^2>0\,,$ and $ T>X$ is:
\begin{align}\label{fromPoin}
Z= \frac{r_+ e^{-r_+ t}}{\sqrt{r^2-r_+^2}}\,,\qquad T= \frac{r\, e^{-r_+ t} \cosh (r_+ \phi )}{\sqrt{r^2-r_+^2}}\,,\qquad X=\frac{r\, e^{-r_+ t} \sinh (r_+ \phi )}{\sqrt{r^2-r_+^2}}\,.
 \end{align}

\section{Generating function of Jacobi polynomials}\label{sec:sum}

In this Appendix we will perform a double sum of multiplication of two Jacobi polynomials which is used in the main text. For that, we use the review on generating functions  in \cite{Sri}; formula (62) in Sec. 2.3 of \cite{Sri} reads
 \begin{align}\label{mastersum2}
\sum_{n}^{\infty}\frac{ n! (-\alpha-\beta)!}{(-\alpha-\beta+n)!} &(x-1)^n(y-1)^{n}t^n P_n^{(\alpha-n,\beta-n)}\left(\frac{x+1}{x-1}\right) P_n^{(\beta-n,\alpha-n)}\left(\frac{y+1}{y-1}\right)\nonumber\\
&= (1-xt)^{\alpha}(1-yt)^{\beta}\,_2F_1\left(-\alpha,-\beta,-\alpha-\beta, \frac{(x-1)(y-1)t}{(1-xt)(1-yt)}\right)\,,
\end{align}
We need also the identity
\be\label{jacobisum0}
\sum_{n}^{\infty}P_n^{(\alpha,\beta)}(x)z^{n}=\frac{2^{\alpha+\beta}}{R(1+R-z)^{\alpha}(1+R+z)^{\beta}}\,,\qquad R\equiv\sqrt{-2 x z+z^2+1}\,.
\ee
Combining the previous formulas, with $y\rightarrow 1/y$, and other basic identities of hypergeometric functions, we can derive the following sum:
%
 \begin{align}\label{mastersum2}
\sqrt{(1-x) (1-y)}^{2 h}r^{h} s^{h}\sum_{k,\bar k=0}^{\infty}\frac{ k! (2 h+\bar k-1)!}{\bar k! (2 h+k-1)!} r^{k} s^{\bar k}({x y})^{\frac{\bar k-k}{2}}& P_k^{(\bar k -k,2 h-1)}(1-2x) P_k^{(\bar k-k,2 h-1)}(1-2y)\nonumber\\
&= \frac{\left(\sqrt{\sigma ^2-1}+\sigma \right)^{1-2 h}}{2 \sqrt{\sigma ^2-1}}\,
\end{align}
where $\sigma$ is defined as
\be\label{geodsum}
\sigma\equiv\frac{-\sqrt{x} \sqrt{y} (r+s)+r s+1}{2 \sqrt{1-x} \sqrt{1-y} \sqrt{r s}}\,.
\ee
For the examples worked out in section \ref{sec:btzads}, $\sigma$ is directly related to the geodesic distance between two endpoints.

\section{Inner product with quasi-normal modes eigenfunctions}\label{sec:quasi}

In this appendix we explore what will happen if in \eqref{innergen}, given by 
\be\label{innergen1}
\langle  U(x_f)  | U(x_i)\rangle=\sum_{k,\bar k}^{\infty} \Phi_{k,\bar k}(s_f)  \Phi^*_{k,\bar k}(x_i)\,.
 \ee
we replaced (without justification) $\Phi_{k,\bar k}$ the more familiar quasi-normal modes for the BTZ black hole.

The quasi-normal modes are defined as the fields in black hole geometries that are purely ingoing at the horizon, and that vanish at infinity. For the BTZ black hole, solutions to $\square^2 \Phi= m^2 \Phi$ with these conditions are found in \cite{Birmingham:2001hc}, imposing separability in its variables:
\be\label{quasinormalsol}
\Phi^{\text{QNM}}(x)=e^{-i \omega t}e^{i l \phi}  \left(\frac{r_+}{r}\right)^{2h} \left(1-\frac{r_+^2}{r^2}\right)^{-\frac{i\omega}{2r_+}} \,_2 F_1\left(h+\frac{i}{2r_+}(l-\omega)\,,h-\frac{i}{2r_+}(l+\omega)\,,2h,\frac{r_+^2}{r^2}\right)\,
\ee
where we have considered the non-rotating case ($r_-=0$), and that the mass of the scalar field is related to the conformal dimension as $h=\frac{1}{2}(1+\sqrt{1+m^2})$. 
The vanishing boundary condition gives the left and right quasi-normal modes:
\be\label{modesBTZ}
\omega_{\pm}=\pm l-2ir_+(n+h)\,.
\ee
%
Using the positive root in \eqref{modesBTZ}, and defining $l=ir_+(k-\bar k)$:
\be\label{quasinormalsolJac}
\Phi^{\text{QNM}}_{k,\bar k}(x)= C_{k,\bar k} e^{-r_+(2ht +k x^++\bar k x^-)} \left(\frac{r^2}{r_+^2}-1\right)^{-h} \left(1-\frac{r_+^2}{r^2}\right)^{\frac{k-\bar k}{2}} \,P_k^{(\bar k-k,2h-1)}\left(\frac{r_+^2+r^2}{r_+^2-r^2}\right)\,
\ee
We have named the previous field $\Phi_{k,\bar k}$ by analogy with the global case, but it does not follow \eqref{raiseAdSfield} for the BTZ differential operators in \eqref{gendifBTZ}. 
\\
\\
Inspired by the global case, we will compute the overlap of two states \eqref{changeb} in the bulk. Evaluating \eqref{innergen1} with \eqref{quasinormalsolJac} gives
\begin{align}\label{sumintBTZ}
\langle  U(x_f)  | U(x_i)\rangle=&\sum_{k,\bar k}^{\infty} \Phi^{\text{QNM}}_{k,\bar k}(s_f)  \Phi^{*\,\text{QNM}}_{k,\bar k}(x_i)\\
&=\sum_{k,\bar k=0}^{\infty}e^{-r_+h(\Delta x^++\Delta x^-)}e^{-r_+( k \Delta x^++ \bar k \Delta x^-)}{\frac{k! (2 h+\bar k-1)!}{\bar k! (2 h+k-1)!}}\left(1-\frac{r_+^2}{r_i^2}\right)^{ \frac{k- \bar k}{2}}\left(1-\frac{r_+^2}{r_f^2}\right)^{ \frac{k- \bar k}{2}}
\nonumber\\
&\times  \left(\frac{r_i^2}{r_+^2}-1\right)^{-h}\left(\frac{r_f^2}{r_+^2}-1\right)^{-h}P_{k}^{(\bar k- k,\, 2h -1)}\left(\frac{r_+^2+r_f^2}{r_+^2-r_f^2}\right)P_{ k}^{(\bar k- k,\, 2h -1)}\left(\frac{r_+^2+r_i^2}{r_+^2-r_i^2}\right)\,.
\nonumber
\end{align}
Using again \eqref{mastersum2}, this time with $x=\left(1-\frac{r_+^2}{r_i^2}\right)^{-1}$, $y=\left(1-\frac{r_+^2}{r_f^2}\right)^{-1}$, $r=e^{-r_+\Delta x^-}$, and $s=e^{-r_+\Delta x^+}$, we see that the result for the results is as well \eqref{resultglobal}, with the geodesic length for the BTZ in \eqref{geodlengthr}.

\section{Integral kernels in CFT representation} \label{app:CFTkernel} 

In this Appendix we describe the mapping between a general $|U\rangle$ state at the boundary in Poincar\'e coordinates and a CFT smearing kernel \eqref{gensmear} in Euclidean signature. Though our interpretation is different, the manipulations here are mathematically very similar to those in e.g. \cite{Goto:2017olq}.
 
We parametrize the group element in terms of three parameters $(\sig, w, \bw)$ as
\be
U_1 = e^{-w_1\frac{i}{2}(-2 L_0 - L_1 - L_{-1})}e^{\sig_1(L_{-1}-L_{1})}e^{+\bw_1\frac{i}{2}(-2L_0 + (L_1 + L_{-1}))}~. \label{Uparamapp} 
\ee
By acting on the Ishibashi state we can rotate it, where the splitting into $G$ and $\bG$ is arbitrary and was picked in this way for later convenience:  
\be
G(e^{-w_1\frac{i}{2}(-2 L_0 - L_1 - L_{-1})}e^{\sig_1(L_{-1}-L_{1})})\bG(e^{-\bw_1\frac{i}{2}(-2L_0 + (L_1 + L_{-1}))})|\Sig_{\rm Ish}\rangle = |U\rangle~.
\ee
We now want to realize the $\slt$ generators geometrically in terms of differential operators acting on $\mathbb{R}^2$. We note that this assignment of generators to operators is not fixed by the algebra alone, as conjugation by any $\slt$ element (or an outer automorphism such as $\Sig_{\rm Ish}$) will leave the algebra invariant. The assignment is instead fixed by the boundary behavior of the gauge connection chosen to be the AdS$_3$ connection; for the choice \eqref{ads3conneuc} the assignment is:
\begin{align}
\p & =  \frac{i}{2}\le(2 \ell_0 + \ell_1 + \ell_{-1}\ri)~, \qquad z \p = -\ha(\ell_1 - \ell_{-1}) ~, \qquad z^2 \p = \frac{i}{2}(2\ell_0 - \ell_1 - \ell_{-1}))~, \\
\bar{\p} & = \frac{i}{2}\le(2 \bell_0 - \bell_{1} - \bell_{-1}\ri)~, \qquad \bar{z} \bar{\p} = \ha\le(\bell_{1} - \bell_{-1}\ri)~, \qquad \bar{z}^2 \bar{\p} = \frac{i}{2}\le(2 \bell_0 + \bell_1 + \bell_{-1}\ri)~. 
\end{align} 
Thus the operation we want to realize is
\be
\sO^{(0)}_{U}(z_1, \bz_1) = e^{w \p_{z_1}}e^{2 \mu z_1 \p_{z_1}}e^{\bw \bar{\p}_{z_1}} K(0,z_1,\bz_1)[\sO]~.
\ee
We will now understand how these operators act on the integral kernel. Clearly two of them are just translations in $z_1$ and $\bar{z_1}$: 
\begin{align}
\exp\le(w \p_1\ri) K(\rho, z_1, \bz_1)[\sO] = K(\rho, z_1+w, \bz_1)[\sO]~, \cr
\exp\le(\bw \bar{\p}_1\ri) K(\rho, z_1, \bz_1)[\sO] = K(\rho, z_1, \bz_1 + \bw)[\sO]~, 
\end{align}
where we use the notation from \eqref{Knot}. The more interesting one is the dilatation, which acts by rescaling $z_1$ (note: actually $z_1$, not the second argument of $K$) by a factor of $e^{2\sig}$:
\be
\exp\le(2\mu z_1 \p_1\ri) K(\rho, z_1, \bz_1) = K(\rho, e^{2\mu} z_1, \bz_1)[\sO]~.
\ee
However, due to the form of the integral kernel, we have the following relation:
\be
K(\rho, e^{2\mu} z_1, \bz_1) = K (\rho + \mu, z_1, \bz_1)~.
\ee
(where this is now a relation that works for the arguments of $K$). To see this, note that
\begin{align}
&\int dz d \bz \le(\frac{e^{-2\rho} - (z - e^{2\mu} z_1) (\bz - \bz_1)}{e^{-\rho}}\ri)^{2h-2} \sO(i z, i\bz) \cr&\qquad = \int dz' d \bz \le(\frac{e^{-2\rho'} - (z' - z_1) (\bz - \bz_1)}{e^{-\rho'}}\ri)^{2h-2} \sO(i z', i\bz)~,
\end{align}
where $\rho' = \rho + \mu$, $z' = e^{-2\mu} z$ and we used the scaling property of $\sO(\lambda z, \bz) = \lambda^{-h} \sO(z, \bz)$. Thus we conclude that 
\be
\exp\le(2\mu z_1 \p_1\ri) K(\rho, z_1 + w, \bz_1) = K(\rho + \mu, z_1 + e^{-2\mu} w, \bz_1)[\sO]~.
\ee
We now construct the desired object:
\begin{align}
\sO^{(0)}_{U}(z_1, \bz_1) & = e^{w \p_{z_1}}e^{-2 \mu z_1 \p_{z_1}}e^{\bw \bar{\p}_{z_1}} K(0,z_1,\bz_1)[\sO] \cr
& = e^{w \p_{z_1}}e^{2 \mu z_1 \p_{z_1}}K(0,z_1,\bz_1+\bw)[\sO] \cr
& = e^{w \p_{z_1}}K(\mu,z_1,\bz_1+\bw)[\sO] \cr
& = K(\mu, z_1 + w, \bz_1 + \bw)[\sO]~.
\end{align}
Now we finally need to act with the overall compensating $\rho$ dilatation:
\begin{align} 
\sO^{(\rho)}_{U}(z_1, \bz_1) & = e^{\rho z\p_{z_1}} e^{\rho \bz \p_{\bz_1}}K(\mu, z_1 + w, \bz_1 + w)[\sO] \cr
& =  K(\mu, e^{\rho}z_1 + w, e^{\rho}\bz_1 + w)[\sO] \cr
& = K(\rho+\mu, z_1 + e^{-\rho} w, \bz_1 + e^{-\rho} w)[\sO]~.
\end{align}
The final relation is thus
\be
\sO^{(\rho)}_{U}(z_1, \bz_1) = K(\rho+\mu, z_1 + e^{-\rho} w, \bz_1 + e^{-\rho} w)[\sO]~,
\ee
which is \eqref{gensmear} in the text. 

\bibliographystyle{JHEP-2}
\bibliography{HigherSpin}

\end{document}